\documentclass[
nofootinbib,
floatfix,
superscriptaddress,
amsmath,amssymb,
twocolumn]{revtex4-1}

\usepackage{anysize}

\usepackage{amsfonts}
\usepackage{amssymb}
\usepackage{graphics}
\usepackage{epsfig}
\usepackage{graphicx}
\usepackage{epsfig}
\usepackage{amssymb}
\usepackage{amsfonts}
\usepackage{amsmath}
\usepackage{xcolor}
\usepackage{float}
\usepackage{enumerate}
\usepackage{slashed}
\usepackage{hyperref}
\hypersetup{colorlinks=true,linkcolor=blue,citecolor=blue}

\usepackage{bbold}

\usepackage{braket}

\usepackage{tikz}

\numberwithin{equation}{section}

\newcommand {\beq} {\begin{equation}}
\newcommand {\eeq} {\end{equation}}
\newcommand{\bea}{\begin{eqnarray}}
\newcommand{\eea}{\end{eqnarray}}
\newcommand{\bit}{\begin{itemize}}
\newcommand{\eit}{\end{itemize}}
\def\nl{\nonumber \\}
\def\Tr{{\rm Tr}}
\newcommand{\tmop}[1]{\ensuremath{\operatorname{#1}}}

\def\a{\alpha}

\def\b{\beta}

\def\l{\lambda}

\def\s{\sigma}

\def\p{\partial}

\def\le{\left(}
\def\ri{\right)}

\def\beq{\begin{equation}}
\def\eeq{\end{equation}}

\def\Ad{{\rm Ad}}

\newenvironment{sistema}%
{\left\lbrace\begin{array}{@{}l@{}}}%
{\end{array}\right.}

\def\1{ \mathbb{1}}
\def\o{ \otimes}

\begin{document}

\title{Geometry of quantum complexity }
\author{Roberto  Auzzi}
\email{ roberto.auzzi@unicatt.it}
\affiliation{Dipartimento di Matematica e Fisica,  Universit\`a Cattolica del Sacro Cuore, Via Musei 41,  25121 Brescia, Italy,  }
\affiliation{ INFN Sezione di Perugia,  Via A. Pascoli, 06123 Perugia, Italy, }
\author{ Stefano Baiguera}
\email{stefano.baiguera@nbi.ku.dk}
\affiliation{ The Niels Bohr Institute, University of Copenhagen, Blegdamsvej 17,   DK-2100 Copenhagen \O, Denmark }
\author{ G. Bruno De Luca}
\email{ gbdeluca@stanford.edu}
\affiliation{ Stanford Institute for Theoretical Physics, Stanford University, Stanford, CA 94306 }
\author{ Andrea Legramandi}
\email{  andrea.legramandi@swansea.ac.uk}
\affiliation{Department of Physics, Swansea University, Swansea SA2 8PP, United Kingdom  }
\author{Giuseppe Nardelli}
\email{giuseppe.nardelli@unicatt.it}
\affiliation{Dipartimento di Matematica e Fisica,  Universit\`a Cattolica del Sacro Cuore, Via Musei 41,  25121 Brescia, Italy,  }
\affiliation{TIFPA - INFN, c/o Dipartimento di Fisica, Universit\`a di Trento,  38123 Povo (TN), Italy }
 \author{ Nicol\`o Zenoni}
\email{nicolo.zenoni@unicatt.it}
\affiliation{Dipartimento di Matematica e Fisica,  Universit\`a Cattolica del Sacro Cuore, Via Musei 41,  25121 Brescia, Italy,  }
\affiliation{ INFN Sezione di Perugia,  Via A. Pascoli, 06123 Perugia, Italy, }
\affiliation{ Instituut voor Theoretische Fysica, KU Leuven, Celestijnenlaan 200D, B-3001 Leuven, Belgium}

\begin{abstract}

  Computational complexity is a quantum information concept that
  recently has found applications in the holographic
  understanding of the black hole interior.
  We consider quantum computational complexity for $n$ qubits 
  using Nielsen's geometrical approach. In the definition of complexity
  there is a big amount of arbitrariness due to the choice of the penalty
  factors, which parameterize the  cost of the elementary computational gates.
   In order to reproduce desired features in holography, such as ergodicity and
  exponential maximal complexity for large number of qubits $n$,
  negative curvatures are required.
  With the simplest choice of penalties, this is achieved at the price of singular sectional curvatures
  in the large $n$ limit.
  We investigate a choice of penalties 
  in which we can obtain negative curvatures
  in a smooth way.
  We also analyze the relation between operator and state complexities,
  framing the discussion with the language of Riemannian submersions.
  This provides a direct relation between geodesics and curvatures in
  the unitaries and the states spaces, which we also exploit to give a closed-form expression
  for the metric on the states in terms of the one for the operators. 
  Finally, we study conjugate points for a large number of qubits in the unitary
  space and we provide a strong indication that maximal complexity
  scales exponentially with the number of qubits in a
  certain regime of the penalties space.

\end{abstract}

\maketitle


\section{Introduction}

An important problem in theoretical quantum computation
is to determine the best quantum circuit  to implement
a desired unitary transformation.
In general, this might be a challenging question.
Moreover, it would be nice to have better theoretical tools 
to prove if a quantum computation problem has or not an efficient solution.
The concept of quantum computational complexity has been introduced 
to answer these questions.
Complexity itself is defined in a rather heuristic 
way as the minimal number of computational gates required to 
build a given unitary operator with some tolerance. 
In order to improve the quantitative understanding,
a geometrical approach to computational complexity in quantum mechanics
was introduced in \cite{Nielsen1}  and further studied in
\cite{Nielsen2,Nielsen3,Nielsen5,Nielsen-Dowling}.
The basic idea is to introduce a Riemannian metric in the space of unitary operators
acting on a given number of qubits, which quantifies how hard it
is to implement a given quantum computational task.
The distance induced by the metric in the space of unitary operators
  is used as a measure of the  complexity 
of the quantum operation.

An additional motivation to study  complexity arises from the desire 
of understanding the physics of the black hole interior
 \cite{Susskind:2014rva,Stanford:2014jda,Susskind:2014moa,Susskind:2015toa,Susskind:2018pmk}.
 Quantum information theory already provided us with many insights 
along the road to understand quantum aspects of gravity.
This is especially powerful in the framework of AdS/CFT.
The concept of entanglement entropy has a natural dual in terms
of area of extremal surfaces \cite{Ryu:2006bv}.
Recently, such a geometric realisation of entanglement led us to a better understanding of the Page curve \cite{Page:1993wv}
for an evaporating black hole, see e.g. \cite{Almheiri:2019psf,Penington:2019npb,Almheiri:2019hni}.

It is natural to conjecture that other features of holographic 
spacetime are encoded in other quantum information quantities, such as complexity.
In the context of AdS/CFT correspondence, the growth of computational complexity 
was proposed as the boundary dual of the growth of the size of the 
Einstein-Rosen bridge connecting the left and the right sides
of an eternal black hole in AdS.
Two main holographic duals for complexity were
proposed:
\begin{itemize}
\item the complexity=volume (CV) conjecture relates complexity to the volume
of an extremal slice anchored to the boundary \cite{Susskind:2014rva,Stanford:2014jda,Susskind:2014moa};
\item  the complexity=action (CA) conjecture relates complexity to the action computed
in the Wheeler-DeWitt patch \cite{Brown:2015bva,Brown:2015lvg}. 
\end{itemize}
Holographic complexity was recently studied in a large variety
of settings, see e.g. \cite{Lehner:2016vdi,Cai:2016xho,Chapman:2016hwi,Carmi:2017jqz,
Braccia:2019xxi,Barbon:2015ria,Auzzi:2018zdu,Auzzi:2018pbc}.
Another promising generalisation  is provided 
by subregion complexity \cite{Alishahiha:2015rta,Carmi:2016wjl,Ben-Ami:2016qex,Abt:2017pmf,Agon:2018zso,Alishahiha:2018lfv,
Caceres:2018blh,Auzzi:2019fnp,Chen:2018mcc,Auzzi:2019mah,Auzzi:2019vyh,Caceres:2019pgf,Hernandez:2020nem,DiGiulio:2020hlz}.
The appropriate notion of complexity in quantum field theory,
dual to these holographic quantities, is still an open problem.
One of the most promising and challenging
approaches is to generalise Nielsen's geometric method to quantum field theory,
see e.g. \cite{Jefferson:2017sdb,Chapman:2017rqy,Hashimoto:2017fga,Camargo:2018eof,Magan:2018nmu,
Caputa:2018kdj,Bueno:2019ajd,Chapman:2018hou, Khan:2018rzm, Hackl:2018ptj, Doroudiani:2019llj,
Balasubramanian:2019wgd,Flory:2020eot,Flory:2020dja}.

A conjecture about the generic time evolution of complexity has been proposed in \cite{Susskind:2015toa}.
In this picture, at early times complexity  grows linearly
for a period that is exponential in the number of qubits $n$.
This initial phase is called the complexity ramp.
At time $t \propto e^n$ it reaches its maximum value
and then it flattens for a very long time $t \propto e^{e^n} $,
doubly exponential in $n$ (this is called the complexity plateau). 
 After this very long time,
quantum recurrence can bring back the system 
to sub-exponential values with non-negligible probability.
This picture, if confirmed, would give us interesting insights on the 
quantum history of black holes.
 For instance, white holes could be thought of
as the gravity duals of a phase of decreasing complexity
due to quantum recurrence.

The geometrical approach by Nielsen is an interesting
direction to put the definition of complexity on firmer 
grounds.
There is an important order zero property that complexity 
must satisfy in order to fit the expectations in \cite{Susskind:2015toa}:
 in the limit of large number of qubits $n$, the maximal complexity should
scale exponentially with $n$.

A full understanding of complexity is still
an open problem already in quantum mechanics.
In particular, there are many ways to define geometric
computational complexity. Riemannian geometry 
is just a possibility. It could be that
Finsler geometry is more appropriate to investigate complexity,
both for quantum computer science 
\cite{Nielsen1} and in the holographic case \cite{Chapman:2018hou}.
Even in the more traditional paradigm of Riemannian geometry, there is 
a lot of ambiguity in defining complexity. Part of it comes from the choice of the penalty
factors for the Hermitian generators of the unitary transformations,
which implement the physical concept that some operations can
be harder than others to perform in a quantum circuit.
The simplest possibility would be to choose a uniform penalty factor,
independent of the number of qubits entangled  by the given 
quantum operation. However this brings to
a maximal allowed complexity which does not scale exponentially with
the number of qubits \cite{Nielsen1} and so it does not match our
expectations.
It was suggested in \cite{Nielsen1} that Finsler metrics with uniform penalty factors
or Riemannian metrics with non-uniform penalties may instead
give an exponential complexity in some regions of the parameter space.

An interesting toy model for many desired features
of  complexity geometry was proposed in
 \cite{Brown:2016wib}, considering geodesics in 
 a compact 2-dimensional space with negative curvature.
 In particular, it was argued that negative curvature gives 
 an interesting crossover between $L^2$ norm at small distances
 and an effective $L^1$ norm at large distances. This allows us to remain in the framework of
 Riemannian geometry, which is easier to deal with  compared  to Finsler geometry.
 
 Another desirable property of complexity metric is the ergodicity of geodesics, which is important
 to apply thermodynamical arguments to complexity evolution \cite{Brown:2017jil,Bernamonti:2019zyy,Bernamonti:2020bcf}.
 Ergodicity in this context refers to the general idea that
 the trajectory of a generic state along a geodesic will eventually visit
 all the allowed portions of the unitary space.
 There are classical mathematical results (see e.g. \cite{anosov})
 showing that the geodesic flow
 on a manifold with all negative sectional curvatures is ergodic.
 The complexity metric with uniform penalty factors
 is positively curved in all the directions and does not have an ergodic 
 geodesic flow. The introduction of non-uniform penalty factors
can make some of the sectional curvatures negative \cite{Nielsen-Dowling},
but not all of them.
If the negative contribution dominates, we expect that the geodesic
motion is still ergodic.

Let us denote with $w$, which we will refer to as the  \emph{weight}, the number of qubits which are simultaneously 
entangled by a given generator.
In \cite{Nielsen-Dowling}, the following choice of penalty
factors was studied in detail for systems of $n$ qubits:
\bea
q(w) &=& 1 \, , \qquad w \leq 2 \, , \nl
q(w) &=& q \, , \qquad w >2  \, .
\label{draconian-q-intro}
\eea
In order to get negative scalar curvature, a penalty factor $q$
of order $4^n$ is needed. This brings to a singular limit where
the negative scalar curvature is dominated by a few negative 
sectional curvatures which diverge in the large $n$ limit.
The penalty choice in (\ref{draconian-q-intro})
was called draconian in \cite{Brown:2017jil}. 
It was argued that this choice is not appropriate to reproduce
black hole properties such as scrambling time and switchback effect \cite{Susskind:2014jwa}.

For this reason, in \cite{Brown:2017jil} a less drastic choice
 of penalty factors was advocated. In this paper we will
 study a variant of this choice:
\beq
q(w)=   \alpha^{w-1} \, ,
\label{progressive-intro}
\eeq
where $\a>1$ is a constant. We will call  the choice (\ref{progressive-intro})
 \emph{progressive penalties}.
In order to understand complexity geometry in an analytic way,
we will propose a large $\a$ limit
in which  complexity geometry
might be studied order by order in the expansion parameter $\a^{-1}$.
The leading order sectional curvatures scale as $\a^0$.
We find closed form for all the curvatures up to
the next-to leading order $\a^{-1}$.

As recently emphasised in \cite{Brown:2019whu}, two different but strongly related
definitions of complexity can be considered for quantum systems:
\begin{itemize}
\item
Unitary complexity quantifies how hard it is to build some unitary operators.
It was physically motivated by the problem of quantum circuit computational complexity
\cite{Nielsen1,Nielsen2,Nielsen3,Nielsen5,Nielsen-Dowling}.
\item
 State complexity quantifies how hard it is to build a unitary
transformation which transforms the reference state to the target state
\cite{Susskind:2014rva,Stanford:2014jda,Susskind:2014moa,Susskind:2018pmk}.
This is the most natural way to apply the notion of complexity to holography.
\end{itemize}
For $n$ qubits, the unitary complexity metric is defined on 
the group manifold $SU(2^n)$ and it is a homogeneous but not isotropic metric.
In particular, homogeneity tells us that scalar quantities (such as curvature)
are constant. The state complexity metric instead is defined on 
$\mathbb{CP}^{2^n-1}$ and it is neither isotropic nor homogeneus.
The number of dimensions is smaller than in the unitary metric, but the geometrical
structure is more complicated, because this space is not homogeneous
and the scalar curvature is not constant.
In this paper we point out that the relation between unitary and states
complexity is a particular case of Riemannian submersion  \cite{oneill}.
For this reason, geodesics on the state space are determined by
just projecting a class of geodesics on the unitary space, the horizontal ones \cite{oneill-geodesics}.
Moreover, the curvatures in the state space
can be obtained from the curvatures in the unitary
space by O'Neill's formula \cite{oneill}. 

Complexity is determined (both in unitary and state spaces)
as the length of the shortest geodesic
which connects two given points. 
Given a geodesic starting from an initial point $P$, there exists
another point along the geodesic where it begins to fail to be the minimal one.
This is called the cut point of the geodesic. The cut locus of a given point $P$
is defined as the set of all the cut points  of the geodesics starting from $P$.
For unitaries complexity, the metric is homogeneus and then it is enough 
to study the cut locus at the identity.
In general, finding the cut locus is a complicated problem.
A useful approach is to consider conjugate points which,  roughly speaking,
are the points of the manifold that can be joined by a continuous
1-parameter family of geodesics.
From a general result in geometry, we know that a given geodesic
fails to be the minimising one after its first conjugate point.
The converse is not true: a geodesic may stop to be minimising 
well before a conjugate point is reached. 
In this paper we study conjugate points
of complexity metric both for one and for a large number of qubits.
From this analysis,
we find an evidence that maximal complexity
scales exponentially with $n$
in the progressive model for large $\a$.

The paper is organised as follows.
In section \ref{sect:unitary} we review some results of \cite{Nielsen-Dowling}
for the complexity geometry  in the unitary space for an arbitrary number of qubits
and we derive a useful explicit formula for sectional curvatures.
In section \ref{sect:few-qubits} we briefly discuss some few qubits examples.
In section \ref{sect:many-qubits} we consider the situation of a large number of qubits $n$:
after a brief review of the draconian case, we study the progressive choice 
of penalties (\ref{progressive-intro}). In section \ref{sect:states-submersion} we discuss state complexity
and we point out the relevance of the Riemannian submersion, which relates
the geometry of the states to the one of the unitaries.
We also derive a closed-form expression for the state metric.
In section \ref{sect:conjugate points} we study the conjugate points in the unitary space
of a simple class of geodesics, given by the exponential
of the generators which are eigenvalues of the penalty matrix.
We conclude in section \ref{sect:conclusions}.
Technical details and examples are deferred to appendices.


\section{Unitary complexity}
\label{sect:unitary}

We will first review several useful results 
 about the geometry of unitary complexity, following \cite{Nielsen-Dowling}.
We will consider the space of unitary operators acting on a $n$ qubits system,
which 
is $SU(2^n)$.
The tangent vector at a generic point $U_0$ can be specified in terms of a traceless 
Hermitian generator $H$, which is the tangent to the curve 
\beq
U(t)=e^{-i H t} U_0 \, ,
\label{ACCA}
\eeq
 evaluated at $t=0$.

For a generic curve $U(t)$ in the space of unitaries determined by the Schr\"odinger equation
$\dot{U} (t) =-i H(t) U(t)$, we can define in general a complexity norm using a suitable Riemannian metric:
\beq
l= \int dt  \langle H(t), H(t) \rangle^{1/2} \, .
\eeq
In our application, we will consider $\langle  \dots \rangle$ to be a positive-definite inner product
 independent of the group point $U$. Such a metric can be therefore defined at the origin of
  the group manifold and it can be mapped to every point of the manifold using right-translations.
   This metric is usually called a right-invariant metric \cite{Milnor:1976pu,Arnold} and can be defined starting from 
  a given scalar product at the origin:
 \beq
\langle  H, K \rangle = \frac{\Tr \, \left[ H \mathcal{G}(K)  \right]}{2^n} \, .
\label{Goperator}
\eeq
Here $\mathcal{G}$ is a positive-definite operator on the space of unitaries, 
i.e. a superoperator. This terminology is common in the quantum information literature.

\subsection{Comments on the choice of basis}

We work with the basis defined by generalised Pauli matrices
 $\sigma$, which are nothing but the tensor products of $n$ matrices, 
 each of which can be either a $SU \le 2 \ri$ Pauli matrix
  $\sigma_i$ ($i = 1 \, , 2 \, , 3$) or the identity $\mathbb{1}_2$. 
  We define the \textit{weight} $w \le \sigma \ri$  as the number of $SU(2)$ 
Pauli matrices involved in the tensor product $\s$.
We will consider only diagonal metrics in our basis, i.e. $\mathcal{G}(\sigma)=q_\sigma \sigma$,  so that the inner-product \eqref{Goperator} reads
\beq
\langle  \sigma, \tau \rangle =  q_\s  \delta_{\s \tau}\, ,
\label{DIAG}
\eeq
and we denote by  $q_{\sigma}$ the
 penalty factor for the generator $\s$
  normalized as 
$ \Tr \le \sigma^2  \ri = 2^n . $
 We call the choice $q_\s=1$ the unpenalised choice.

The generalised Pauli matrices have a useful property:
if we choose two elements of the basis, they either commute or anti-commute.
 In the one qubit case this follows directly from the Pauli matrices algebra
 and it can be easily generalised to the $n$ qubits case. In particular,
let us  consider the product $\tau \s$ of two generalised Pauli matrices. Then we have
\beq
\label{comm gen pauli}
\s \, \tau = \le -1 \ri^{l} \tau \, \s \, ,
\eeq
where $l$ is the number of the corresponding entries in the tensor products 
in $\tau$ and in $\s$ involving different Pauli matrices.

It is useful  to count the number of generalised Pauli matrices anticommuting with a given $\sigma$. 
If $\sigma = \mathbb{1}$, trivially there are no operators anticommuting with it. 
If $\sigma \neq \mathbb{1}$, a generalised Pauli matrix $\rho$ anticommutes with it 
under the condition that there is an odd number $l$ of corresponding entries in the tensor
 products in $\sigma$ and $\rho$ involving different Pauli matrices.
  Let us suppose that $\sigma$ has weight $w$ (its tensor product contains $w$ Pauli matrices).
   Then, we necessarily have $0 \leq l \leq w$. Among the $n$ entries of the tensor product in $\rho$,
   the $n-w$ entries in correspondence with the identity  $\mathbb{1}_2$ in $\s$
   can arbitrarily  be any matrix in the basis $(\mathbb{1}_2,\sigma_i)$ indifferently. 
    Thus we have $4^{n-w}$ choices for such entries.     
For the remaining $w$ entries of $\rho$, we have $\binom{w}{l}$ choices for the $l$ 
positions of the unequal Pauli matrices. Once this is fixed, there is a further $2^w$ 
degeneracy of choices.
   Summarizing, the number of generalised Pauli matrices $\rho$ anticommuting with $\sigma$ is
\bea
\label{number-anticommu}
4^{n-w} \sum_{l \, {\rm odd} = 1}^{w} \binom{w}{l} \, 2^w  
= \frac{4^n}{2} \, .
\eea
It is remarkable that  
  the number of $\rho$ anticommuting with a given 
  $\sigma \neq \mathbb{1}$ does not depend on the weight of $\sigma$.

The commutator of two elements of the basis 
 (if not vanishing)  is proportional to another element of the basis, 
 because the two products in the commutator give the opposite matrix ($l$ is odd). Given two non-commuting elements of the basis $\s$ and $\tau$, we define 
 $q_{[\s,\tau]}$ as the penalty of their commutator;
  if $[\s,\tau]=0$  we set by definition  $q_{[\s,\tau]}=1$.  

\subsection{Connection and geodesic equation}

Let us now derive an expression for the Levi-Civita connection $\nabla$ compatible with the metric \eqref{Goperator}. This is given by the Koszul formula \cite{petersen}, which, thanks to the fact that the inner product can be computed at the identity (and therefore is constant in a suitable basis), simplifies to
\begin{equation}
\label{eq:koszul}
-2 i \langle \nabla_X Y, Z \rangle = \langle [X , Y ], Z \rangle+\langle [Z , X ], Y \rangle -\langle [Y , Z ], X \rangle 
\end{equation}
where $X,Y,Z$ are right-invariant fields interpreted as Hermitian matrices at the origin.
Eq.  \eqref{eq:koszul} allows us to define
\beq
\nabla_X Y=\frac{i}{2}\le [X,Y]+
 \mathcal{G}^{-1} ( [X, \mathcal{G}(Y)]+[Y, \mathcal{G}(X)] ) \ri \, .
 \label{deri-hamiltonian-rep}
\eeq

Setting $Y=X$ in eq. (\ref{deri-hamiltonian-rep}),  we obtain  the geodesic equation,
which is nothing but the Euler-Arnold\footnote{Recent applications of the Euler-Arnold equations in relation 
to complexity were found in \cite{Flory:2020eot, Flory:2020dja}.} equation \cite{Arnold}:
\beq
\dot{X}+ i \mathcal{G}^{-1} \le [X, \mathcal{G}(X) ] \ri = 0\, .
\label{geogeo}
\eeq
In general we expect that geodesics have an intricate behaviour.
 Eq. (\ref{geogeo}) shows that there exists a simple class of geodesics,
given by the exponential of an eigenvector
of the penalty operator $\mathcal{G}$.
We will call the geodesics which are exponential
of such eigenvectors "exponential geodesics".
We study the behaviour of their conjugate points in section \ref{sect:conjugate points}.


\subsection{Riemann tensor}

Let us now specialize the discussion to SU$(2^n)$ using Pauli matrices $\rho \, , \sigma \, , \tau \, , \mu$, which can be viewed as  right-invariant frame fields. The curvature tensor is \cite{Nielsen-Dowling}
\beq
R_{\rho \sigma \tau \mu} = \braket{\nabla_{\rho} \tau \, , \nabla_{\sigma} \mu}
 - \braket{\nabla_{\sigma} \tau \, , \nabla_{\rho} \mu} - \braket{\nabla_{i \left[ \rho \, , \sigma \right]} \tau \, , \mu} \, .
\eeq 
Using eq. (\ref{deri-hamiltonian-rep}), we find:
\beq
\label{definition c}
\nabla_{\sigma} \tau = i \, c_{\sigma \, , \tau} \left[ \sigma \, , \tau \right] \, , \qquad 
c_{\sigma \, , \tau} \equiv \frac{1}{2} \le 1 + \frac{q_{\tau} - q_{\sigma}}{q_{\left[ \sigma \, , \tau \right]}} \ri \, .
\eeq
The Riemann tensor  is given by the expression:
\bea
\label{riemann}
 R_{\rho \sigma \tau \mu} &=& c_{\rho \, , \tau} \, c_{\sigma \, , \mu} \braket{i \left[ \rho \, , \tau \right] \, , i \left[ \sigma \, , \mu \right]} \nl
&&- c_{\sigma \, , \tau} \, c_{\rho \, , \mu} \braket{i \left[ \sigma \, , \tau \right] \, , i \left[ \rho \, , \mu \right]} \nl
 && - c_{\left[ \rho \, , \sigma \right] \, , \tau} \, \braket{i \left[ i \left[ \rho \, , \sigma \right] \, , \tau \right] \, , \mu} \, . \nl
\eea
Since eq. (\ref{riemann}) depends just on commutators,
 the Riemann curvature of a subgroup of unitaries 
does not depend on the metric data outside this subgroup.
For example, complexity on 
a one qubit subgroup depends just on penalties of generators acting
on that particular subgroup. 
An important result \cite{Nielsen-Dowling}
 is that the component $R_{\rho \sigma \tau \mu}$ vanishes unless 
the product of the corresponding generalised Pauli matrices $\rho \sigma \tau \mu$ 
is  proportional to the identity.


\subsection{Sectional curvatures}

The sectional curvature is defined as half of the scalar curvature of a 2-dimensional submanifold  with 
tangent space specified by the directions $(\rho, \s)$.
The general expression for the sectional curvature of the plane 
determined by the vectors $(v,w)$ is \cite{john-lee}
\beq
K(v,w)=\frac{R_{\a \b \gamma \delta} v^\a w^\b w^\gamma v^\delta }
{(v_\a v^\a)(w_\b w^\b) -(v_\a w^\a)^2 } \, .
\label{definition-K}
\eeq
The quantity $K(v,w)$ depends just on the plane which is defined by $(v,w)$
and does not depend on their normalization.
The sectional curvature is a non-linear object and it is 
a non trivial function of the orientation of the plane;
in general, in order to determine $K$ on an arbitrary plane
it is not enough to determine it on the planes defined by couples
of vectors on an orthogonal basis.

The generalised Pauli matrices are orthogonal but not normalized, see eq. (\ref{DIAG}).
The sectional curvature in the plane spanned by two generalised Pauli matrices is
\beq
K \le \rho \, , \sigma \ri = \frac{R_{ \rho \sigma  \sigma \rho }}{q_{\rho} \, q_{\sigma}} \, .
\eeq
From eq. (\ref{riemann}) we find
\bea
R_{\rho \sigma \sigma \rho}  &=& c_{\rho \, , \sigma} \, c_{\sigma \, , \rho} \braket{i \left[ \rho \, , \sigma \right] 
\, , i \left[ \sigma \, , \rho \right]} \nl
 && - c_{\left[ \rho \, , \sigma \right] \, , \sigma}
 \, \braket{i \left[ i \left[ \rho \, , \sigma \right] \, , \sigma \right] \, , \rho} \, . \nl
\eea
This vanishes if $\rho$ and $\sigma$ commute. 
Instead, in the case of anticommuting $\rho$ and $\sigma$,
a direct calculation gives:
\bea
\label{obs1}
&& \braket{i \left[ \rho \, , \sigma \right] \, , i \left[ \rho \, , \sigma \right]} = 4 \, q_{\left[ \rho \, , \sigma \right]} \, ,  \nl
&& \braket{i \left[ i \left[ \rho \, , \sigma \right] \, , \sigma \right] \, , \rho}  = -4 \, q_{\rho} \, ,
\eea
where in both the relations we repeatedly used the fact that $\rho$ and $\sigma$ anticommute. 
We can also use the property  $q_{\left[ \left[ \rho \, , \sigma \right] \, , \sigma \right]} = q_{\rho}$
to get the sectional curvature
\bea
\label{sectional_curv}
&& K \le \rho \, , \sigma \ri = \, \frac{1}{q_{\rho} \, q_{\sigma}} \times \\
&&  \left[ -3 \, q_{ \left[ \rho \, , \sigma \right]} + 2 \le q_{\rho} 
 + q_{\sigma} \ri + \frac{\le q_{\rho} - q_{\sigma} \ri^2}{q_{ \left[ \rho \, , \sigma \right]}} \right] \, ,
 \nonumber
\eea
which is valid if $[\rho,\sigma] \neq 0$
(otherwise $K \le \rho \, , \sigma \ri=0$).

This formula, which as far as we know is new and not contained in \cite{Nielsen-Dowling},
has interesting consequences.
We see that the only negative contribution
to $K \le \rho \, , \sigma \ri $ comes from $q_{ \left[ \rho \, , \sigma \right]}$:
$K$ can become negative only if the commutator $\left[ \rho \, , \sigma \right]$
has a large enough penalty factor.
In general, we expect that $K$ is positive, unless $q_{ \left[ \rho \, , \sigma \right]}$
is big enough compared to $q_\rho$ and $q_\s$.

One may wonder if it is possible to get negative all the sectional curvatures
of the orthogonal basis. This is not possible, because the sectional curvatures of the one qubit
subspace depend just on the one qubit penalty factors.
In section \ref{1qubit-unitaries} we will show that at least $2$ out of $3$ independent sectional
curvatures are always positive for one qubit. 


\subsection{Ricci tensor and curvature}

Sectional curvatures are related to Ricci tensor and Ricci curvature.
As shown in  \cite{Nielsen-Dowling},  in our basis
the only non-vanishing component of the Ricci tensor
$R_{\s \tau}$ are the diagonal ones, with $\s=\tau$.
Given an orthonormal basis $\lbrace e_k \rbrace $ with $ k=1,\dots ,N$ and such that $e_1=v$, 
we have the following result \cite{john-lee} valid for all Riemannian manifolds:
\beq
R_{\a \b} v^\a v^\b = \sum_{k=2}^N K(v,e_k) \, .
\eeq 
In this way the scalar curvature can be expressed in terms of the sectional curvatures as 
\beq
\label{scalar_sectional}
R   =\sum_{k=1}^N  R_{\a \b} e_k^\a e_k^\b
=\sum_{\s, \, \rho} K \le \rho \, , \sigma \ri\, .
\eeq
It should be emphasised that the sectional curvatures do not
transform linearly as tensors, still their sum reproduces the Ricci scalar.

The sign of sectional curvatures plays a key role in relation to ergodicity \cite{Brown:2017jil}.
Roughly speaking, the geodesic flow is called ergodic if its
typical geodesic will eventually pass nearby to all the allowed
portions of the operator space. The average
of observables along the geodesic trajectory will then
coincide with the average over the manifold of unitaries.
In the context of the motion in the group manifold of unitaries, one can consider the time evolution of two neighboring geodesics intersecting at $t=0$ under infinitesimally close local Hamiltonians. 
In such a case, the deviation between the geodesics is governed by the sectional curvature corresponding to the section containing the two geodesics: if the sign is positive as in the standard inner product metric, then the geodesics converge.
On the other hand, an appropriate choice of penalty factors allows to obtain negative sectional curvatures, implying that the geodesics diverge.
The divergence of geodesics is an important requirement for quantum chaos, which in turn requires an ergodic behaviour. 

From a general theorem \cite{anosov}, we know that geodesic flow is ergodic in manifolds whose all sectional curvatures are negative.
This result is not directly applicable to unitary complexity, because
at least some of the sectional curvatures in the one qubit directions are always positive.
Indeed, ergodicity of geodesic is still preserved in some examples
where the curvature is partly negative and partly positive (see e.g. \cite{burns-gerber}).
In general, we expect that the presence of directions 
with mostly negative sectional curvatures
is a strong indication of ergodic behaviour of geodesics.
From eq. (\ref{scalar_sectional}) we know that
the scalar curvature is the sum of all the sectional curvatures 
of an orthogonal basis, and so we expect that negative scalar
curvature $R$ is a detector of ergodicity. Unfortunately, we don't know about
any rigorous mathematical theorem which  relates the sign of $R$
to the ergodicity of geodesics.

In view of the investigation of conjugate points of the geodesics in section \ref{sect:conjugate points}, it is convenient to introduce a specific notation for the diagonal components of the Ricci tensor. 
Using an orthonormal basis $\lbrace u(\sigma) \rbrace$ in the algebra, we define
\bea
&& R_\s=R_{\a \b}u^\a(\s) u^\b (\s)=\frac{R_{\s \s}}{q_\s} \, ,  \nl
&& u^\a (\sigma) u^\b (\sigma) g_{\a \b}=1 \, .
\label{ricci-diagonal-def}
\eea
Using the definitions for the curvature quantities given above, we start by considering in section \ref{sect:few-qubits} the simple cases where the quantum-mechanical system is composed by one or two qubits.
We will extract the sectional curvatures and the Ricci scalar and study their behaviour in relation to various choices of the penalty factors on the generators.
Then we will generalize in section \ref{sect:many-qubits} to the case with many qubits, where we will propose some choices of penalty factors to reproduce expected properties of complexity.


\section{Few qubits examples }
\label{sect:few-qubits}

\subsection{One qubit }
\label{1qubit-unitaries}

\begin{figure*}
\includegraphics[scale=0.52]{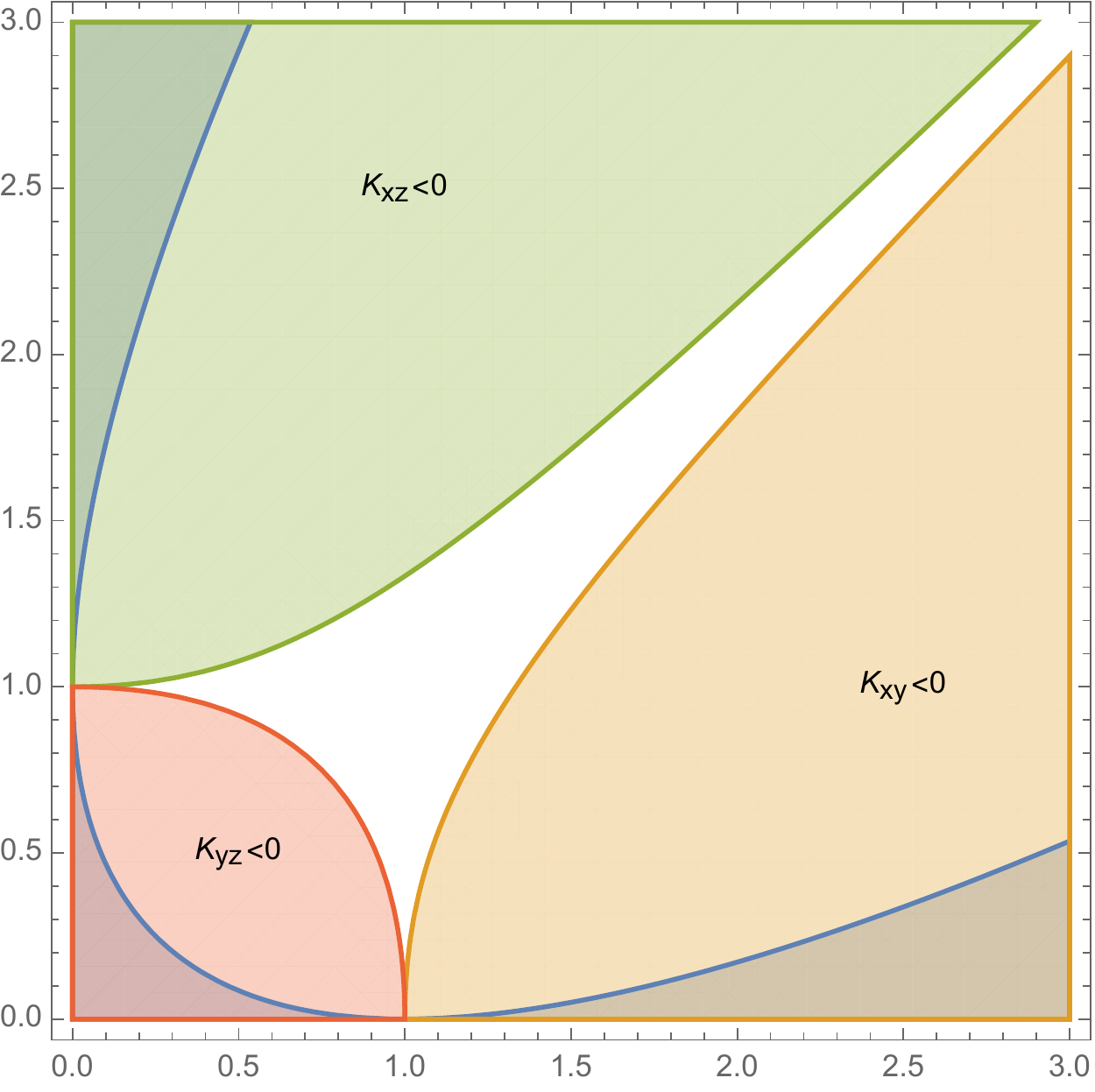} 
\caption{ \label{regions-map-unitary-1qubit}
Regions of negativity of sectional curvatures in the $(P,Q)$ plane.
In the white region all the sectional curvatures are positive.
The blue shaded regions correspond to a negative scalar curvature.}
\end{figure*}

Let us fix the penalty factor for $\s_x$ to $1$ and 
denote the penalty factors for $\s_y$ and $\s_z$ by $Q$ and $P$.
  For $Q=1$,  the metric has a $U(1)$ isotropic
  symmetry which rotates $(\s_x,\s_y)$.
 Applying the results of the previous section, the
  sectional curvatures of the planes selected by our orthonormal basis are:
\bea
K_{x y} &=& \frac{-3 P^2 + 2 P + 2 P Q +Q^2 +1 -2 Q}{P Q} \, , \nl
K_{x z} &=& \frac{-3 Q^2 + 2 Q + 2 P Q +P^2 +1 -2 P}{P Q} \, ,  \nl
K_{y z} &=& \frac{-3  + 2 P + 2  Q +P^2 +Q^2 -2 P Q}{P Q} \, , \nl
\label{sectional-1qubit}
\eea 
  and the scalar curvature is
 \bea
R=- 2 \frac{(Q-P)^2-2 (P+Q)+1}{P Q} \, .
\label{1qubit-R}
\eea
The signs of sectional and scalar curvatures
 are shown in  Fig. \ref{regions-map-unitary-1qubit}.
Note that  two out of the three
sectional curvatures in eq. (\ref{sectional-1qubit})
are positive in all the parameter space.

The sectional curvatures form a non-linear object; these quantities are not enough
to compute the sectional curvature in an arbitrary plane,
which can be found using expressions from the Riemann tensor.
In the one qubit case, we checked that the values 
in eq. (\ref{sectional-1qubit}) correspond for all $P,Q$ to 
 the maxima and minima of the sectional curvature.

Conventionally, we will call the generators with lowest
penalty "easy" generators, and those with highest penalty "hard" generators.
We are interested in limits where 
the maximal complexity becomes large,
in general exponential in the number of states.
So it might seem a contradiction to search
for limits of large complexity in the one qubit 
Hilbert space. This is not necessarily the case:
in order to explore a toy model with large maximal
complexity, one may consider the limit 
where the weight factors $P,Q$ go to infinity.

One of these limits
may be obtained by setting 
\beq
P=1 \, , \qquad Q \rightarrow \infty \, .
\label{1qubit-draconian}
\eeq
 In this case the scalar
and the sectional curvatures diverge:
\bea
&& R=8-2 Q \, , 
\qquad
K_{x y} = 4- 3 Q \, ,  \nl 
&& K_{x z}=K_{y z }= Q \, .
\label{curvatures-1qubit-draconian}
\eea
In general, if we set $P$ constant and we send $Q \rightarrow \infty$,
we do not obtain a smooth limit.

It is also interesting to consider  the limit
 \beq
P=Q \rightarrow \infty \, .
\label{1qubit-moderate}
\eeq
The scalar curvature remains small:
\bea
&& R=\frac{8}{P} - \frac{2}{P^2} \, , \qquad
K_{x y}=K_{x z}=\frac{1}{P^2} \, , \nl
&& K_{y z}=\frac{4}{P}-\frac{3}{P^2} \, .
\label{curvatures-1qubit-moderate}
\eea
In this case all the sectional curvatures are positive and become small.

Another possibility is to consider
 \beq
P=\b Q \rightarrow \infty \, ,
\label{1qubit-moderate-2}
\eeq
with $\b$ constant. At large $P$
we find that the sectional curvatures approach to constants.
For $\b \neq 1$, at large $P$ the scalar curvature is negative,
$R=-2 (\b-1)^2$.

In all these limits the volume of the space
(measured using the complexity metric) goes to infinity.
From the point of view of complexity, 
instead, these limits are very different.
In the case (\ref{1qubit-draconian}) the maximal complexity does not approach
infinity, because the remaining easy generators are enough to build
whatever unitary we want.
Instead, in the cases in eq. (\ref{1qubit-moderate}) 
and (\ref{1qubit-moderate-2}) the maximal complexity goes
to infinity, because the only easy generator at our disposal allows to produce just
a very special class of unitary, i.e. the rotations  along the $x$ axis.

\subsection{Two qubits}

The two qubits case is the simplest environment where we 
can address the question of what happens if one penalizes
operators according to the number of qubits that are 
entangled at the same time.  

We  choose $A$ as penalty factor for the weight $1$ matrices 
and $B$ as penalty factor for the weight $2$ ones.
The non-vanishing sectional curvatures $K(\rho, \s)$
in the orthonormal basis can take three values:
\beq
a=\frac{1}{A} \, , \qquad b=\frac{A}{B^2} \, , \qquad  c=\frac{4 B -3 A}{B^2} \, .
\eeq
The value $a$ arises when $(\rho, \s)$ have both weight $w=1,$ 
the value $c$ when they have both $w=2$ and the value
$b$ if they are generators with different weights.
The multiplicity of each value of the sectional curvatures is:
\beq
N_a=12 \, , \qquad N_b=72 \, , \qquad N_c=36 \, .
\eeq
The scalar curvature is:
\beq
R = -12 \, \frac{3 \, A^2 - 12 \, A B - B^2}{A B^2} \, .
\eeq
Let us specialise $A = 1$ and $B = q$ with $q>1$.
We are penalizing  the weight $2$ matrices (denoted as "hard")
compared to the weight $1$ matrices (denoted as "easy").
The scalar curvature is
\beq
R = 12 \, \frac{-3 + 12 \, q + q^2}{q^2} \, ,
\eeq
which is always positive. 
Note that in this case the structure of the algebra generators is as follows
\bea
\label{e-h-1}
&& \left[ {\rm easy} \, , {\rm easy} \right] = {\rm easy} \, , \qquad
 \left[ {\rm easy} \, , {\rm hard} \right] = {\rm hard} \, , \nl
&&  \left[ {\rm hard} \, , {\rm hard } \right] = {\rm easy} \, ,
\eea
and so it gives rise to positive sectional curvatures, from eq. (\ref{sectional_curv}).
Although such a choice is the most intuitive, it necessarily provides  positive 
curvatures,  see also \cite{Brown:2019whu} for the same conclusion. 
Note that no singularity appears in the curvature if we send $q \rightarrow \infty$.

If we instead set $A=p$ and $B=1$, we are penalizing the weight $1$ matrices and the scalar curvature is 
\beq
R = -12 \,  \frac{3 \, p^2 - 12 \, p -1}{p} \, .
\eeq
Note that in this case the structure of the algebra generators is reversed
\bea
\label{e-h-2}
&& \left[ {\rm hard} \, , {\rm hard } \right] = {\rm hard} \, , \qquad \left[ {\rm easy} \, , {\rm hard} \right] = {\rm easy} \, , \nl
&& \left[ {\rm easy} \, , {\rm easy} \right] = {\rm hard} \, ,
\eea 
and indeed gives negative curvature at large enough $p$,
according to eq. (\ref{sectional_curv}).
This result gives a quantitative explanation of some intuitions discussed in \cite{Brown:2019whu}.

We point out that the abovementioned case is not the only one where such a behaviour occurs. 
In general, when we split the set of generators in two classes, one of which is a maximal subalgebra, the structure of 
commutators \eqref{e-h-1} and \eqref{e-h-2} always arises.


\section{Many qubits}
\label{sect:many-qubits}

We consider quantum systems composed by many qubits,
which is the first step in the direction of a
system with infinite degrees of freedom as it happens in field theory.
In this case it is possible to study the dependence of the curvatures on the number of qubits, in order to understand the assignment of penalty factors that can reproduce physical phenomena like the switchback effect and scrambling.

The idea is to study the time evolution of complexity when the system of interest is subject to a perturbation.
From the holographic point of view, this is usually performed with the introduction of a shock wave very far in the past, in such a way that the scrambling time corresponds to the delay after which the black hole reaches again the equilibrium \cite{Stanford:2014jda}.
From the perspective of quantum circuits, a useful model consists in the evolution of an epidemic \cite{Susskind:2018pmk}. 
If there is a single infected qubit which can interact with all the other ones via a local Hamiltonian, the scrambling time measures the scale after which the infection has involved a large enough number of qubits in order for complexity to reach the value $n,$ corresponding to the number of qubits.

In this context, a related effect is the switchback one, which is a delay in the growth
 of complexity arising from cancellations between multiple shock waves or perturbations.
 Using the toy model introduced in \cite{Brown:2016wib},
it was suggested that, in order to get a satisfying description of
switchback effect and scrambling,  the typical sectional curvatures
should scale as $1/n$ or $1/n^2$ in the large number of qubits limit
(depending on the variant of the model).
For a recent discussion of the switchback effect for low number of qubits,
see \cite{Caginalp:2020tzw}.
Even without restricting to a particular toy model,
the divergence of sectional curvatures in the large number of qubits 
limit gives rise to a singular behaviour that should be avoided.
In this section we will study
 the consequences of various assignments of penalties on the behaviour of curvatures.

Let us consider the case with $n$ qubits,  
equipped with a class of penalty factors which are
functions just of the weight of the generators.
Let us denote the penalty associated to the weight $k$ by  $q_k$.
The number of generalised Pauli matrices with weight $k$ in our basis is given by:
\beq
\mathcal{N}_k=3^k \binom{n}{k} \, .
\label{eq:N_subscript}
\eeq
   Given two generators $(\rho, \sigma)$,
   let us denote respectively by $M$ and $N$ the weights of $\rho$
   and $\sigma$, and by $w$ the weight of the commutator $[\rho,\sigma]$.
   From the analysis given in appendix \ref{counting-appendix},
   we can show that $w$ can take the following values
   \beq
w_r= | M-N | +1 + 2 r \, , 
\label{wr-main}
\eeq
where the integer $r$ has the following range
\beq
0 \leq r \leq \min(M,N) -1 \, .
\eeq

If  two directions in the unitary space do not commute,
the sectional curvatures can be obtained from eq. (\ref{sectional_curv}), i.e.
 \bea
&& K(M,N,r) =
\frac{1}{q_M q_N} \times \nl
 && \! \! \! \! \! \! \! \le 
-3 \, q_{ w_r} +
 2 \le q_{M} + q_{N} \ri 
 + \frac{\le q_{M} - q_{N} \ri^2}{q_{w_r}}
\ri 
\label{values-K}
\eea
where $K(M,N,r)$ denotes the sectional curvature of the plane spanned by generalised Pauli matrices of
 weights $M$ and $N$, whose commutator has weight $w_r,$ given by eq. (\ref{wr-main}).
 We denote by $\mathcal{N}(M,N,r)$
the degeneracy of such sectional curvatures.
We derive an explicit expression for  $\mathcal{N}(M,N,r)$
in appendix \ref{counting-appendix}.

If two directions commute $K(M,N,r) =0$;
given a generalised Pauli matrix, about one half
of the other Pauli matrices in the basis commute
with it, see eq. (\ref{number-anticommu}).
So about one half of the total sectional curvatures 
vanish by construction, independently of the penalty factors.

\subsection{Draconian penalties}

The combination of 1 and 2 qubits operators is universal
and can be used to build an arbitrary operator in $SU(2^n)$ \cite{Nielsen-book}.
This result suggests a somewhat minimal choice of penalty factors, 
studied in detail in \cite{Nielsen-Dowling}
\bea
q_\s &=& 1 \, , \qquad w \leq 2 \, , \nl
q_\s &=& q \, , \qquad w >2  \, .
\label{draconian-q}
\eea
This choice does not distinguish different values of the weight $w>2$
and was called "draconian" in  \cite{Brown:2019whu}.

The sectional curvatures can be found using the general expression in eq. (\ref{sectional_curv}), giving
the values in Table \ref{table1}. 
\begin{table}[h]     
\begin{center}    
\begin{tabular}  {|l|l|l|} 
\hline         & $\left[ \rho \, , \sigma \right] \in \mathcal{P}$ & $\left[ \rho \, , \sigma \right] \in \mathcal{Q}$     \\ \hline
$\rho, \sigma \in \mathcal{P}$  & $K=1$ & $K=4 - 3  q$ \\ \hline      
$\rho, \sigma \in \mathcal{Q}$  & $K=\frac{4 \, q - 3}{q^2}$ & $K= \frac{1}{q}$ \\ \hline      
$\rho \in \mathcal{P}$, $\sigma \in \mathcal{Q} $ & $K=q$ & $K= \frac{1}{q^2}$ \\ \hline      
\end{tabular}   
\caption{\footnotesize 
Values of the non-vanishing 
sectional curvature $K$ for various choices  of $\rho,\s$ in the model with draconian penalty factors.
We denote by $\mathcal{P}$ and $\mathcal{Q}$  the set of generators with $w \leq 2 $ and $w>2$ respectively.
 }     
\label{table1}  
\end{center}
\end{table}

For $q =1$ we recover the case where all the penalty factors are
equal, which corresponds to a bi-invariant metric on $SU(2^n)$.
In this case all the non-vanishing  sectional curvatures are equal and positive.
The interesting region with negative curvature is at large $q$.
So in this limit it makes sense to use the approximation
where only the  sectional curvatures at leading order in $q$
 are considered.

Let us consider the approximation in which we keep just the $\mathcal{O}(q)$
and the $\mathcal{O}(1)$ terms.
In this limit the only non-vanishing sectional curvatures are:
\bea
K(1,1,0) &=& K(2,1,0)=K(2,2,0)=1 \, , \nl
K(3,2,0) &=& q \, , \nl 
 K(2,2,1)&=&4-3 q \, ,
\eea
with multiplicities
\bea
\mathcal{N}(1,1,0) &=& 6 n \, , \nl
 \mathcal{N}(2,1,0) &=& \mathcal{N}(2,2,0)=18 n (n-1) \, , \nl
\mathcal{N}(3,2,0) &=&  \mathcal{N}(2,2,1)=54 n (n-1) (n-2) \, . \nl
\eea
The scalar curvature then is:
\beq
R=-54 n (n-1) (n-2) q +6 n ( 36 n^2 - 99 n + 64 ) \, .
\eeq

This calculation is in agreement with the exact result computed in  \cite{Nielsen-Dowling}
in a different way:
   \begin{widetext}
\beq
\begin{aligned}
R & = - 54 \, q \, n \le n-1 \ri \le n-2 \ri + 6 n \left[ 36 n^2 - 99 n + 64 \right] + \\
 &+ \frac{1}{q} \left[ \frac{4^n}{2} \le 4^n -1 + \frac{3 n \le 3 n -1 \ri}{2} \ri - 6 n \le 45 n^2 - 117 n +74 \ri \right] + \\
& - \frac{1}{q^2} \left[ 3 n \le 3 n -1 \ri \, 4^{n-1} - 6 n \le 3 n -4 \ri \le 6 n -7 \ri \right] \, .
\end{aligned}
\eeq 
   \end{widetext}
In order to get negative curvature, we need $q \propto 4^n$ or larger.
This means that $q$ has to grow exponentially with $n$.
In particular, in this regime the scalar curvature 
is dominated by a small number (polynomial in $n$)
of sectional curvatures whose magnitude grows like $|K| \approx q \approx 4^n$.
This is a singular limit, and, as discussed in \cite{Brown:2017jil},
this brings to some unwanted properties in  the
 scrambling and switchback effect of
black holes complexity.

\subsection{Towards a more sustainable taxation policy}

In \cite{Brown:2017jil} a more moderate 
penalty factors choice was advocated:
\bea
q_\s &=& 1 \, , \qquad w \leq 2 \, , \nl
q_\s &=& c \, 4^{w-2} \, , \qquad w >2  \, , 
\label{moderate}
\eea
where $c$ is an order $1$ constant.
The authors called this choice "moderate",
because sectional curvatures are not as big as 
in the draconian model.
Big curvatures in general are not a desired feature
of complexity geometry, because they are in tension
with the desired properties of scrambling and switchback effect.
The exponential behaviour $q_k \propto 4^k$ in (\ref{moderate})
is suggested by the draconian model: in such a case the behaviour
 $q \propto 4^n$ of penalties is needed 
in order to have negative curvature.
In this section we will consider some variations of this model,
in which $q_k \propto \a^k$ for some appropriate constant $\a$.

The draconian model resembles a flat tax: all the weights
bigger than $2$ are treated the same. 
The  middle-class exponents with $w \approx 3$
and the billionaires with $w \approx n$ pay exactly the same amount of taxes.
The penalty choice in  eq. (\ref{moderate}) goes in the direction of a 
more progressive taxation, because high incomes are taxed progressively.
Still there is a minor 
source of  inequality in  eq. (\ref{moderate}): the very low income guys
at $w=1$ are taxed just the same as the working class at $w=2$.
In order to promote social justice
we are motivated to introduce the following choice of 
penalties (see also \cite{Susskind:2020wwe})
\beq
q_\s=   \alpha^{w-1} \, ,
\label{progressive}
\eeq
which we will call "progressive" penalties.
The scaling as $4^k$ at large $k$ is generalised as $\a^k$.

The model (\ref{progressive}) simplifies  in the 
large $\a$ limit, which can be used as an expansion parameter
for the analytical understanding of the model. In particular, 
from eq. (\ref{sectional_curv}) we can see that
at large $\a$ sectional curvatures scale at most as $\a^0$.
With the choice in eq. (\ref{progressive}), we expect that 
by construction the maximal complexity becomes infinity at fixed $n$
in the limit $\a \rightarrow \infty$, because one qubit
operators cannot produce the most general operators
in the unitary space.
For example, they cannot produce unitaries which
entangle two qubits that were previously unentangled.
Physically, we will be interested in the limit of large but finite $\a$.

Moreover, we can  consider generalisations of this basic model.
In particular, we can generalise the choice in eq. (\ref{progressive}) as
\bea
q_\s &=& 1 \, , \qquad w \leq w_0 \, , \nl
q_\s &=& \a^{w-w_0} \, , \qquad w > w_0 \, ,
\label{moderate-with-threshold}
\eea
with $w_0 \geq 2$.
For $w_0=2$ and $\a=4$, we recover the model studied in
\cite{Brown:2017jil}.
With this choice of penalties, we expect that the maximal 
complexity at fixed $n$ does not diverge for $\a \rightarrow \infty$,
because the combination of 1 and 2 qubits operators is universal
and can be used to build an arbitrary operator in the unitary space.
From  eq. (\ref{sectional_curv}), we can see that
this model has the property that at large $\a$
sectional curvatures scale at most as $\a^{w_0-1}$.
Therefore, the large $\a$ limit  provides a singular geometry,
as the curvature diverges.

\subsection{Progressive penalties}

We computed the curvatures as a power expansion in $\a$,
at the leading order $\a^0$ and at next to leading order $\a^{-1}$.
The cumbersome calculations are deferred to appendix
\ref{progressive-appendix}.

At the leading order in $\a$, the scalar curvature is:
\beq
R = 3 n \le 4^n - 2 \, 7^{n-1} \ri \, .
\label{scalar-curvature-progressive}
\eeq
It is negative for $n \geq 3$ and comes just from two values
of the sectional curvatures:  $K=1$ with multeplicity
$\mathcal{N}_+$ and $K=-3$ with multeplicity $\mathcal{N}_-$, where
\bea
&& \mathcal{N}_+  = 12\ 7^{n-1} n-3\ 2^{2 n+1} n+18 n \, ,
\nl
 && \mathcal{N}_- = \frac{\mathcal{N}_+}{2}- 3n  \, .
\eea
At next to leading order, the correction to the curvature is:
 \beq
 \delta R=\frac{9}{2} n (n-1) \frac{4^n}{\a} \, .
 \eeq

In order to get a feeling on the average sectional curvature,
it is convenient  to divide $R$ by the total number
 of sectional curvatures between couples of elements in the basis,
  which we denote by 
\beq
\eta=(4^n-1)^2 - (4^n-1) \, .
\eeq
The average sectional curvature becomes tiny at large $n$ and $\a$, i.e.
\beq
\bar{K}=\frac{R}{\eta} \approx - \frac{6}{7} n \left( \frac{7}{16} \right)^n
+\frac{1}{\a} \frac{9}{4^n} \frac{n  (n-1)}{2}   \, .
\eeq
We don't have an analytic expression at higher order in $\a$
for the generic $n$ qubits case. However, if $n$ is 
fixed to be some not too large value, we 
can compute the exact result at all orders explictly
 since the sum over the penalties contains a finite  number of terms.

\begin{figure*}[ht]
\includegraphics[scale=0.52]{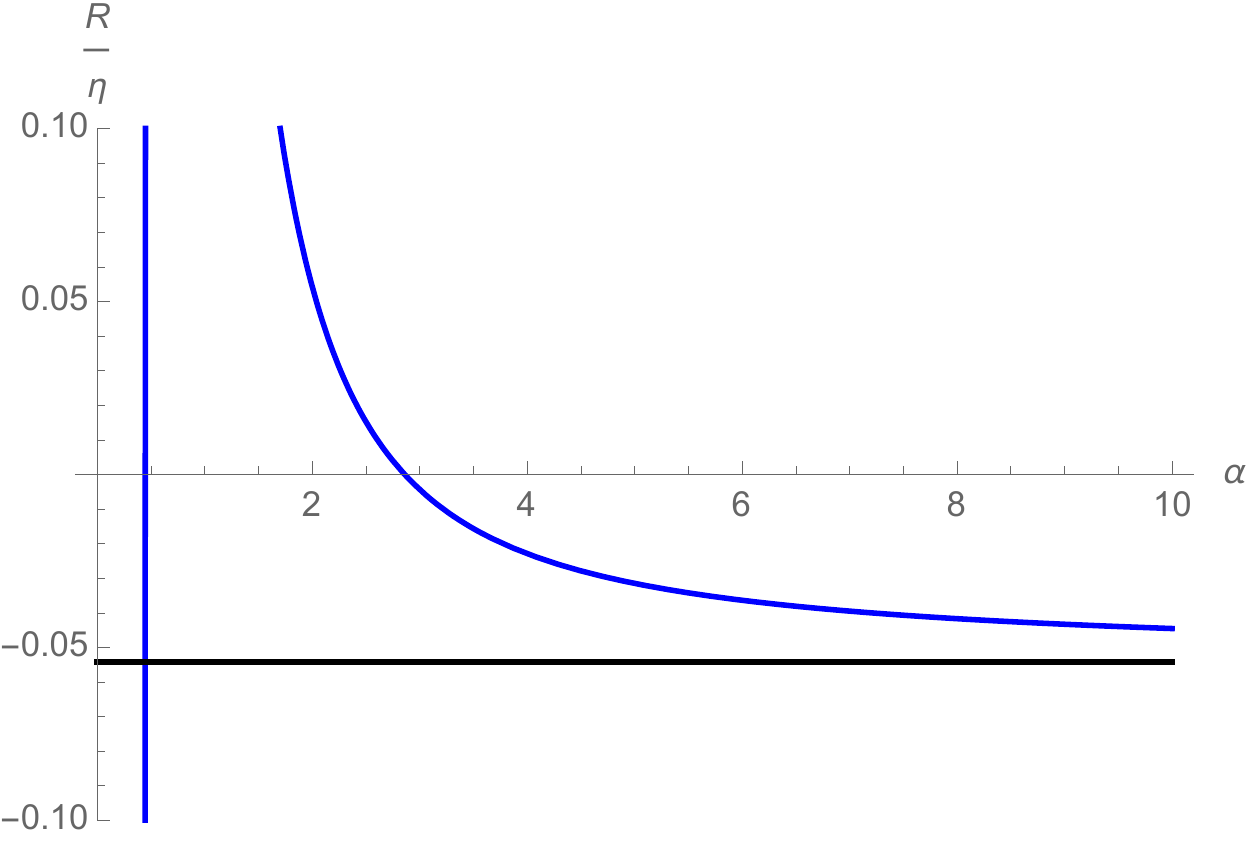} \qquad
\includegraphics[scale=0.52]{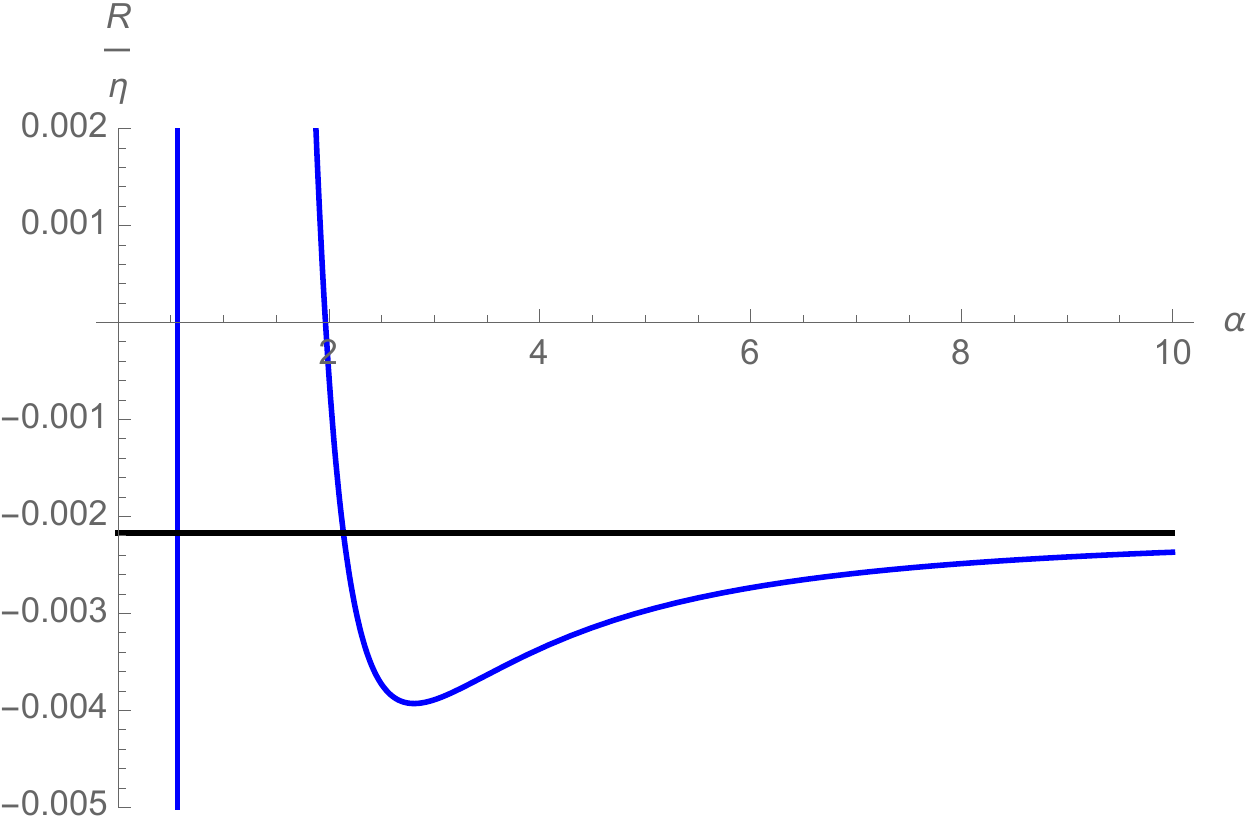} \qquad
\includegraphics[scale=0.52]{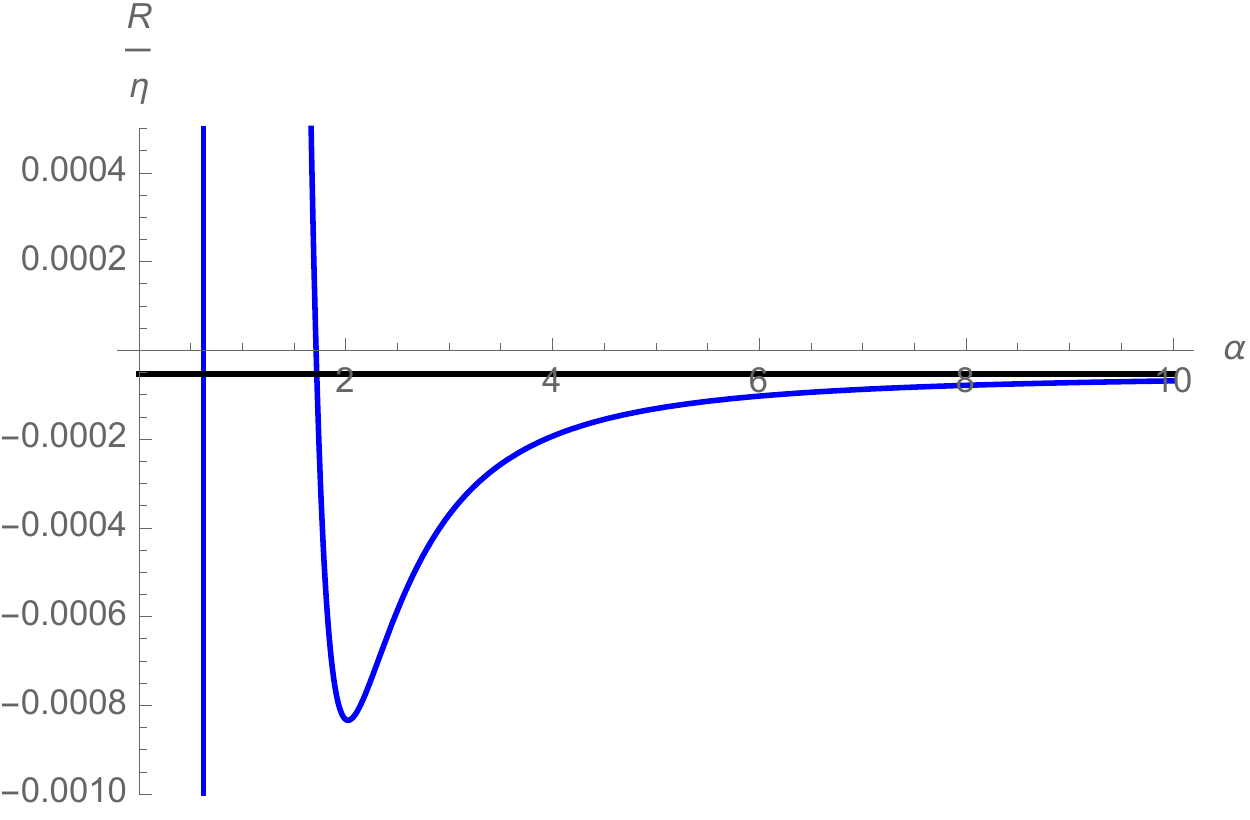} \qquad
\includegraphics[scale=0.52]{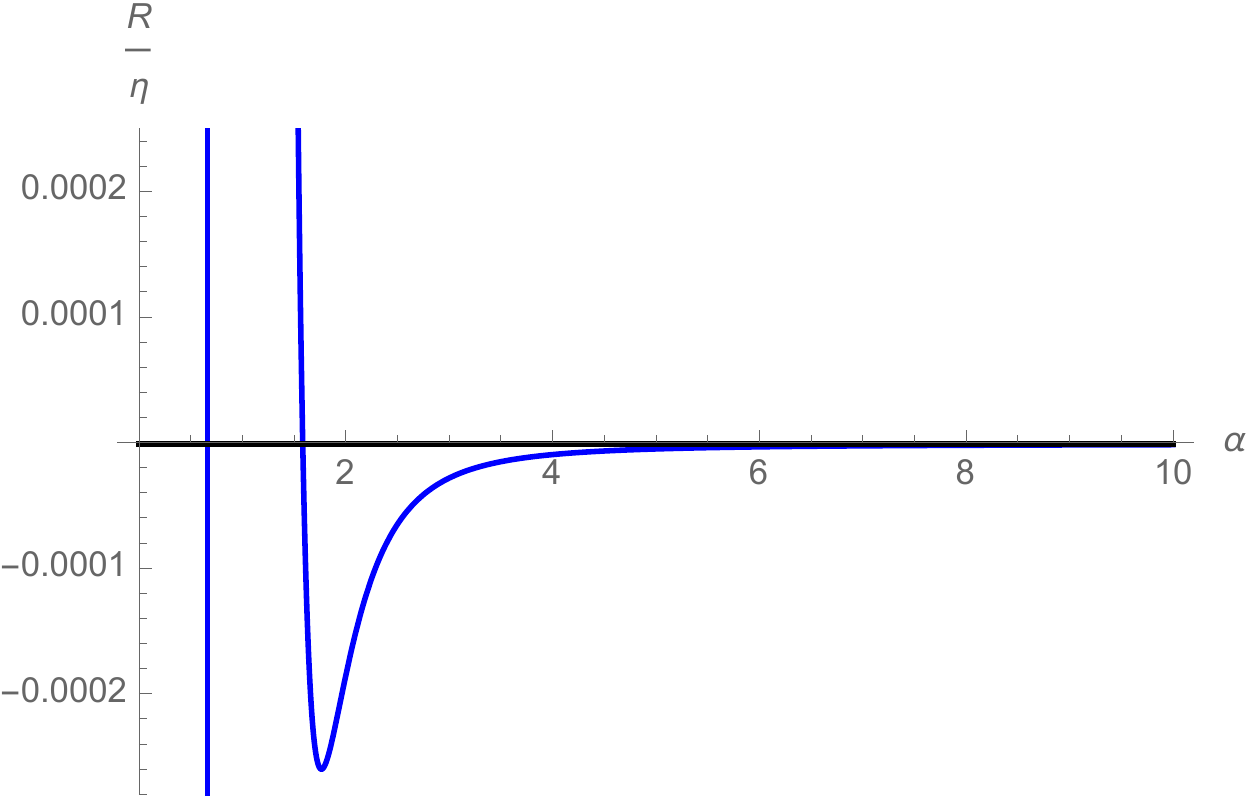} 
\caption{ \label{alpha-fig}
The exact value of $R/\eta$ plotted as a function of $\alpha$ in the case of progressive penalties,
 for $n=5,10,15,20$. The asymptotic value at $\a \rightarrow \infty$ is shown in black.
  The minimum in the picture appears for $n \geq 8$.
  Increasing $n$, the shape of the minimum tends to become more and more steep
  and it is located at a lower value of $\a$.
Note that when $n=20,$ for values $\alpha \geq 4$ 
 the result of the average sectional curvature at $\mathcal{O}(\alpha^0)$ is already very close to the exact result. 
}
\end{figure*}

 
The exact value of the average sectional curvature as a function of $\a$ 
for a few values of $n$ is plotted  in Fig. \ref{alpha-fig}.
Nothing special happens for the value $\a=4$,
which instead plays an important role for the  draconian model.
It is interesting that there is a minimum at finite $\a$.
It turns out that the series expansion in $\a^{-1}$ for $\bar{K}$ is, 
at large $n$, an alternate sign series 
with slow rate of convergence.
For example, in order to get the minimum in the plot for $\bar{K}$ when $n=10$,
we have to expand up to the order $\a^{-5}$.

This choice of penalties
for $\a \rightarrow \infty$ 
has many similarities with the
one qubit case in eq. (\ref{1qubit-moderate-2}),
where $P= \b Q \rightarrow \infty$ with $\beta$ constant 
and different from $1$.
In both limits we expect that the maximal complexity
diverges, and  the sectional curvatures do not.
Also, $R$ approaches a negative constant in both cases.


\section{State complexity and submersions}
\label{sect:states-submersion}

Up to now, we have focused the discussion on the complexity of \emph{unitaries}.
In this section, we bring the attention of our reader to the geometry of the space of \emph{states}.
Geometrically, this space is naturally associated with a quotient of the space of unitaries where all the different unitary transformations that, starting from a given reference state, build the same state (up to a phase) are identified.
The complexity of the state built in this way is then defined to be the minimum of the complexities of all the identifed unitaries.
Requiring that the state complexity is also obtained as a length on the space of states defines a map between two Riemannian manifolds, which turns out to be a \emph{Riemannian submersion}. 
We recall its definition in section \ref{sec:submersions}, and we proceed in the subsequent sections in exploiting known results for Riemannian submersions.

In particular, O'Neill's formula relates the curvature of the space of states to the one of the space of unitaries, providing a lower bound on the curvature on states.
This underlying geometrical structure allows a direct comparison of some class of geodesics, which we explore in section \ref{sect:submersions-geodesics} and \ref{sect:conjugate points}.

\subsection{Submersions}\label{sec:submersions}
For convenience of the reader, in this section we briefly review the concept of Riemannian sumbersions, referring
 to the textbooks \cite{petersen, besse} for more details.

\begin{figure*}
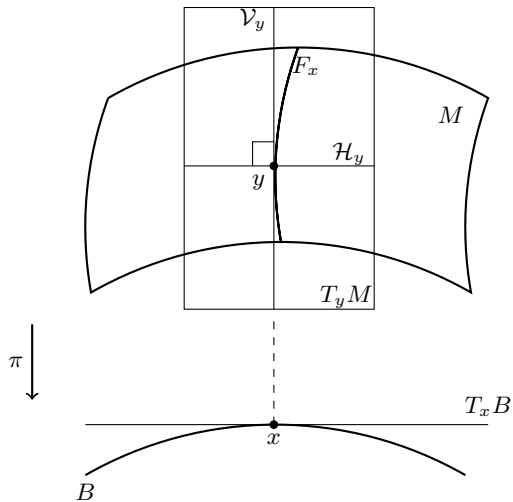

\begin{center}
\tikz{[scale=2] 
\draw[thick] (0,0) arc[start angle=120, end angle=60, radius=5]
arc[start angle=160, end angle=190, radius=5]
arc[start angle=60, end angle=90, radius=5]
arc[start angle=190, end angle=160, radius=5]
arc[start angle=160, end angle=190, radius=5]
arc[start angle=90, end angle=120, radius=5]
arc[start angle=190, end angle=160, radius=5];
\node[below left] at (4.8,0) {$M$};
\node[below right] at (2.3,5-5*sqrt 3/2) {$F_x$};
\draw[thick] (-0.3,-5) arc[start angle=120, end angle=60, radius=5];
\node[below] at (-0.3,-5) {$B$};
\draw (-0.3,-5*sqrt 3/2) -- (5,-5*sqrt 3/2);
\draw[fill] (2.18,-5*sqrt 3/2) circle [radius=.05];
\node[below] at (2.18,-5*sqrt 3/2) {$x$};
\node[above] at (5,-5*sqrt 3/2) {$T_x B$};
\draw[dashed] (2.18,-5*sqrt 3/2)--(2.18,-2.9);
\draw[->,thick] (-1,-3) -- (-1,-4);
\node[left] at (-1,-3.5) {$\pi$};
\draw (1,-2.8) rectangle (3.5,1.2);
\draw (2.18,1.2)--(2.18,-2.8);
\draw (1,-0.9)--(3.5,-0.9);
\draw (1.9,-0.9) |- (2.18,-0.58);
\node[above left] at (3.5,-1) {$\mathcal{H}_y$};
\node[below left] at (2.2,1.3) {$\mathcal{V}_y$};
\draw[fill] (2.18,-0.9) circle [radius=.05];
\node[below left] at (2.18,-0.9) {$y$};
\node[above left] at (3.6,-2.9) {$T_y M$};}
\end{center}
\caption{ \label{submersion}
A reproduction of a depiction of a submersion from \cite{besse}.}
\end{figure*}

Let us consider two Riemannian manifolds $(M, g_{\a \b})$ with dimension $m$
and $(B,h_{\a \b})$ with dimension $b < m$
and a smooth map $\pi: M \rightarrow B$ with surjective differential $d\pi$.
$d\pi$ is a map $d\pi: TM \to TB$ , that for any $y\in M$ induces a linear map between the vector spaces $T_y M$ and $T_{x} B$, where $x = \pi(y)$.
This map has maximal rank, and thus a kernel of dimension $f = m - b$.
We will call $\mathcal{V}_y = \text{ker}(d\pi_y)$ the \emph{vertical space at $y$.} 
Its orthogonal complement in $T_y M$, induced by the metric $g$, is called the \emph{horizontal space at $y$} and denoted by $\mathcal{H}_y$.
For the submersion to be Riemannian,  $\mathcal{H}_y$ has to be identified with $ T_x B$ in an isometric way, in other words
\begin{equation}
  g(X,Y)=h(d\pi(X),d\pi(Y)),\qquad \forall X,Y \in \mathcal{H}_y \;.
\end{equation}
A pictorial depiction is shown in Figure \ref{submersion}.

Quotients of manifolds by an isometric group action provide interesting examples of submersion
(see for example the textbooks \cite{petersen, besse}). Let $M$ be a Riemannian manifold 
and $G$ be a closed subgroup of the isometry group of $M$, and denote by $\pi$
the projection from $M$ to the quotient space $B = M/G$.
This defines a natural metric on $B$ such that $\pi$ is a Riemannian submersion \cite{petersen}.

In the following sections, we make use of this construction to understand properties of the space of states from the complexity of unitaries.


\subsection{Submersions and complexity geometry}
\label{submersion-explicit}

Let us apply the notion of submersion to the complexity geometry.
We take $M=SU(2^n) $ with a right-invariant metric (the unitary space)
and $G$ as the subgroup of the isometries of $M$
 which leaves the reference state invariant up to a phase.
More precisely, we consider
a unitary $U$ which generates the state $| \psi \rangle$
starting from the reference state  $| \psi_0 \rangle$
\beq
U | \psi_0 \rangle =| \psi \rangle \, .
\eeq
We call \emph{unbroken subgroup} the subgroup of $SU(2^n$) that fixes the reference state up to a phase
\beq
V | \psi_0 \rangle = e^{i \phi}| \psi_0 \rangle \, .
\eeq
Such a $V$ is an element of $SU(2^n-1)\times U(1)$.
Thus, up to a phase, both $U' \equiv U V$ and $U$ prepare the same state $|\psi\rangle$: 
\beq
U'| \psi_0 \rangle = e^{i\phi}| \psi \rangle \, ,\qquad \implies \quad U' \sim U \, .
\eeq
Therefore we have a map from the unitary space to the quotient $B$ defined as
\beq
\pi : \qquad SU(2^n) \rightarrow B
\label{map-pi}
\eeq
where
\beq
B=\mathbb{CP}^{2^n-1}=\frac{SU(2^n)}{SU(2^n-1) \times U(1)} \, .
\eeq
This map is an isometric submersion, as we are going to prove writing it explicitly
in a specific coordinate system.

In order to make contact with section \ref{sect:unitary}, we take a diagonal penalty matrix in the basis of the generalised Pauli matrices, see eq. (\ref{DIAG}), with the property
\beq
\langle \s_r, \s_s \rangle = q_r \, \delta_{r s}=q_r \,  \frac{1}{2^n} \Tr (\s_r \s_s) \, .
\eeq
For the states metric it is more convenient to do a change of basis.
We can identify a basis for  broken generators $ \rho_k$ and unbroken ones $\tau_a$:
\bea
&&\omega_l=( \rho_k,  \tau_a) \, , \qquad
 1 \leq k \leq 2 (K-1) \, ,
 \nl
&&  1 \leq a \leq  (K-1)^2 \, ,  \qquad K=2^n \, ,
  \label{omegarhotau}
\eea
with normalization
\beq
\Tr \, (\omega_l \, \omega_m) = \delta_{m l} \, .
\eeq
We can express
\bea
&& \s_r =\sum_l \omega_l \, \Tr \le \omega_l \s_r \ri \, , \nl
&& \omega_l=\frac{1}{2^n} \sum_r \s_r \, \Tr \le \omega_l \s_r \ri \, .
\eea
Then we can find the penalty scalar product in the basis $\omega^k$:
\beq
M_{l m} = \langle \omega_l, \omega_m \rangle =\frac{1}{2^{2 n }}   \, \sum_r q_r \Tr \le \omega_l \s_r \ri   \Tr \le \omega_m \s_r\ri \, .
\eeq
This discussion also applies to the case where $M_{lm}$ is a generic symmetric matrix.
Let us introduce the following notation for the exponential of broken and unbroken generators
\beq
U_\theta =e^{ i  \theta_k  \rho_k }  \, , \qquad  V_\l=e^{ i  \l_a  \tau_a } \, ,
\eeq
 where the variables $\theta_k$ denote the coordinates in the state space and $\lambda_a$ are the additional coordinates that define the unitary space.
A generic element of $SU(K)$ can be written as
$
U=U_\theta V_\l  .
$
Then we can compute
\bea
dU U^\dagger &=& (dU_\theta V_\l + U_\theta dV_\l ) V_\l^\dagger  U_\theta^\dagger \nl
&=& dU_\theta  U_\theta^\dagger +  U_\theta dV_\l V_\l^\dagger  U_\theta^\dagger 
 \, ,
\eea
where
\beq
d U_\theta=\frac{\p U_\theta}{\p \theta_k} \, d \theta_k \, , \qquad
d V_\l=\frac{\p V _\l}{\p \l_j} \, d \l_j \, .
\eeq
In this way the right-invariant forms defined on $\mathrm{SU}(K)$ are given by 
\bea
X_r&=& -i \, \Tr (dU \, U^\dagger \, \omega_r)  \nl
&=& -i (\Ad_{U_\theta^\dagger})_{r s} \Tr  \left\{ (U_\theta^\dagger dU_\theta+ dV_\l V_\l^\dagger)  \omega_s \right\} \, .
\nonumber
\eea
where we have used the adjoint action
\beq
U_\theta^\dagger \omega_r U_\theta = (\Ad_{U_\theta^\dagger})_{rs} \omega_s \, .
\eeq
We can now write the metric in the unitary space as
\beq
ds^2=M_{r s } X_r X_s=\tilde{M}_{l m} (u_l + v_l) (u_m +v_m) \, ,
\label{uuu-metric}
\eeq
where
\bea
\tilde{M}_{l m} &=& M_{r s} (\Ad_{U_\theta^\dagger})_{r l}   (\Ad_{U_\theta^\dagger})_{s m} \, ,
\nl 
u_s &=& -i \, \Tr  \left\{  U_\theta^\dagger dU_\theta  \omega_s \right\} \, , \nl
v_s &=&  -i  \, \Tr  \left\{  dV_\l V_\l^\dagger   \omega_s \right\} \, ,
\label{eq:right_left_inv_forms}
\eea
in such a way
 that
$ \tilde{M}_{l m}$ depends just on $\theta_k$,
  $u_s$ contains just $(\theta_k, d \theta_k)$
and  $v_s$ contains just  $(\l_a, d \l_a)$.

Now it is convenient to split the indices in $\omega_r$ in 
indices corresponding to broken and unbroken generators, as in eq. (\ref{omegarhotau}).
We have that $v_i=0$ for $i$ corresponding to a broken index.
Then we can write the unitary metric eq. (\ref{uuu-metric}) as
\bea
ds^2 &=&
\left(
\begin{array}{cc}
 u_i  &  u_a +v_a  \\
\end{array}
\right)
\left(
\begin{array}{cc}
 \tilde{M}_{ij} &  \tilde{M}_{ib}  \\
   \tilde{M}_{aj}  &  \tilde{M}_{ab}   \\
\end{array}
\right)
\left(
\begin{array}{c}
 u_j  \\  u_b +v_b  
\end{array}
\right) \nl
&=&  (\tilde{M}_{ij}- \tilde{M}_{ i c}  \tilde{M}_{ca}^{-1} 
 \tilde{M}_{ a j} )  u_i  u_j  +\tilde{M}_{ab} f_a f_b \, ,  
    \nonumber \\
    \label{EMME-sub}
\eea
where we introduced 
\beq
 f_a = v_a + u_a +  \tilde{M}_{ad}^{-1} \tilde{M}_{ d j} u_j \, \, .
\eeq
The problem of finding the minimal infinitesimal operator which synthesizes
the state of coordinates $\theta_k+d \theta_k$ from the state with coordinates $\theta_k$ is
then solved by the equation $f_a=0$, because the term $\tilde{M}_{ab} f_a f_b$
in eq. (\ref{EMME-sub}) is positive-definite.
This  construction generalises the result in \cite{Brown:2019whu}
to arbitrary number of qubits.

We can then identify the metric on the state space $B$ as
\beq
ds^2_S= (\tilde{M}_{ij}- \tilde{M}_{ i c}  \tilde{M}_{ca}^{-1} 
 \tilde{M}_{ a j} )  u_i  u_j \, .
 \label{metric-states-general}
\eeq
We explicitly checked that the metric in the space of states $\mathbb{CP}^1$
 for a single qubit coincides with the result found in  \cite{Brown:2019whu}.
 In appendix \ref{sect-qudit_3states} we will see how to apply this result to qutrits.

From eq. (\ref{EMME-sub}),  it follows that the 
projection map $\pi$ from $M$ to $B$
\beq
\pi: (\theta_k, \l_j) \mapsto (\theta_k)
\eeq
is a Riemannian submersion,
where $\pi^{-1}(\theta_k)$ is parametrised by $\l_k,$ for fixed $\theta_k.$
The explicit expression for the horizontal
spaces at arbitrary $\theta_k$ is given by $f_a (X)=0$ for
any generic vector $X$ in the tangent space.

  
\subsection{Submersions and curvature}
 
We can use O'Neill's formula \cite{oneill}  to relate 
the sectional curvatures of states $K_S$ to the ones of unitaries $K$:
\beq
K_S (\tilde{h}_1, \tilde{h}_2)=K(h_1, h_2) 
+\frac{3}{4}  \frac{| \mathcal{V} ([h_1,h_2]) |^2}{ |h_1|^2 |h_2|^2 -\langle h_1, h_2 \rangle^2 } \, ,
\label{oneill}
\eeq
where $\mathcal{V}$ is the projector on the vertical subspace, 
$\langle \dots \rangle$ is the scalar product
 from the metric of the manifold $M$, $|\dots|$
 is the norm induced by the scalar product,
$\tilde{h}_k=d \pi(h_k)$ are vectors fields in the state space,
$h_k$ are horizontal fields in the unitary space,
 $[h_1,h_2]$ is the commutator of the  vector fields
 in the unitary space.

This expression shows that the sectional curvature of a plane
 in the space of states can be always expressed 
 as sectional curvature of an appropriate plane in the unitary space
 plus a positive definite contribution coming from the commutator of horizontal vectors.
 It can be used to compute the curvatures in the state space without 
 even knowing its metric.
As an illustrative example, we apply eq. (\ref{oneill})
to the one qubit case in appendix \ref{1qubit-appe}.

\subsection{Submersions and geodesics}
\label{sect:submersions-geodesics}
 
 The relation between geodesics in $B$ and geodesics in $M$
 for generic submersions was studied in  \cite{oneill-geodesics}.
An important result is that if a geodesic in $M$  is horizontal at some point, it remains horizontal.
Then the projection by $\pi$ of an horizontal geodesic
 is a geodesic in the space of states $B$.
 As a general result, we have that
 for submersions from complete manifolds $M$ as our unitary space,
 every geodesic of $B$ can be built as the projection
 of a horizontal geodesic in $M$.
 It is important to stress that the projection of 
 a geodesic which is not horizontal in general does not
 provide a geodesic on $B$.
 
 We know from eq.  (\ref{geogeo})  that the exponential 
 of an eigenvector of the penalty matrix $\mathcal{G}$ is a geodesic
 in the unitary space. Combining with the previous result,
 the exponential of an eigenstate of $\mathcal{G}$ which is also
 perpendicular to the unbroken subgroup at the origin,
 gives a geodesic in the state space $B$.
 This property provides us a simple class of geodesics in
 some particular situations.
 In the 1-qubit case, this is studied in section \ref{sect-Geodesics_one_qubit}.
 
 Let us instead consider the 2-qubits case with penalties depending just on the
 weights.  Taking as reference state $| 00 \rangle $, the unbroken subgroup
 is generated by the following generators:
 \bea
&&\1 \o \s_z \, , \qquad  \s_z  \o \1 \, ,  \qquad \s_z \o \s_z \, , 
\nl
 &&\s_x \o (\1-\s_z ) \, , \qquad  \s_y  \o (\1-\s_z ) \, , \nl
&&  (\1-\s_z ) \o \s_x  \, , \qquad   (\1-\s_z ) \o \s_y \, , 
\nl
&& \s_x \o \s_y -\s_y \o \s_x \, , \nl
&& \s_x \o \s_x +\s_y \o \s_y \, .
\label{unbrok-2}
 \eea
The orthogonal complement to this space is generated by:
 \bea
 && \s_x \o (\1+\a \, \s_z ) \, , \qquad  \s_y  \o (\1+\a \,  \s_z ) \, , \nl
&& (\1+\a \, \s_z ) \o \s_x  \, , \qquad   (\1 +\a \, \s_z ) \o \s_y \, ,  \nl 
&& S^-_2= \s_x \o \s_y + \s_y \o \s_x \, , \nl
&&   S^+_2= \s_x \o \s_x - \s_y \o \s_y  \, ,
\label{brok-2}
 \eea
 where $\a$ is a coefficient\footnote{The precise value is completely irrelevant for the following discussion.} dependent on the penalty factors, chosen to ensure orthogonality
 with unbroken generators in eq. (\ref{unbrok-2}).
 Note that just the last two generators $S^{\pm}_2$ in eq. (\ref{brok-2})
  have a definite weight $w=2$,
 and so just these two operators generate exponential horizontal geodesics.
 
We can generalise this arguments to $n$ qubits as follows.
Let us take as reference state $|00 \dots 0 \rangle$.
Let us consider the action of a infinitesimal transformation on this state,
 with $w=n$  and which contains just $\s_x$ and $\s_y$ entries in the tensor product.
This operator will rotate the state as
\beq
|00 \dots 0 \rangle \rightarrow  |00 \dots 0 \rangle + \epsilon  |11 \dots 1 \rangle \, ,
\eeq
where $\epsilon $ is an infinitesimal complex number.
This sector of operators contain $2^n$ generators;
out of this set, a vector space of dimension $2^n-2$ operators is unbroken.
So, in the $w=n$ sector which contain just tensor products of $\s_x$
and $\s_y$ we can always find a 
broken dimension $2$ subspace which is orthogonal to
the vertical space

Let us build these generators explicitly.
We introduce
\beq
A^n_s= \frac{1}{\binom{n}{s}} \sum_{(k_1,\dots, k_n)} \s_{k_1} \o \dots \o \s_{k_n} \, 
\eeq
where the sum runs over all the permutations
 $(k_1,\dots, k_n)$ which contain  $s$ generators $\s_y$ and $n-s$ generators $\s_x$.
 Then the two generators
\beq
S_n^+=\sum_{0 \leq k \leq n}^{k \,\,\, {\rm even}} i^{k} \, A^n_{k} \, , \qquad
S_n^-=\sum_{0 \leq k \leq n}^{k \,\,\, {\rm odd}} i^{k+1} \, A^n_{k} \, .
\label{Spm}
\eeq
are both broken by the reference state and
 orthogonal to all the unbroken $w=n$ generalised Pauli matrices
which contain just $\s_x$ and $\s_y$ in the tensor product.
This construction generalises to $n$ qubits the two
operators in the last line of eq. (\ref{brok-2}).

Then we can look for other generators orthogonal to the vertical space.
We can consider a generalised Pauli matrix of the form
$S_{n-1}^\pm \o (\1 + \a_1 \s_z)$ with the coefficient chosen in such a way
that it is orthogonal to  $S_{n-1}^\pm \o (\1 -\s_z)$.
This involves a linear combination of weight $n$ and $n-1$ generators
and in general one can find $2 \binom{n}{1}$ such operators.
One can iterate the construction, looking 
for generators of the form 
\beq 
S_{n-s}^\pm \o (\1+\a_s \s_z)^s \, ,
\label{Sn-s}
\eeq
and determine $\a_s$ in such a way that (\ref{Sn-s})
 is orthogonal to the unbroken operators
\beq
S_{n-s}^\pm \o (\1- \s_z) \o \1^b \o \s_z^c \, , 
\eeq
where $b,c$ are some integer numbers.
  For each integer $s$, the operators in (\ref{Sn-s}) 
are linear combinations of weight $w$ generators with 
\beq
n-s \leq w \leq n \, .
\label{s-proiettabile}
\eeq
There are  $2 \binom{n}{s}$ of such operators, with $1 \leq s \leq n $.
In this way one can build all the $2^n-1$ horizontal vectors
in the unitary space, which project to the $\mathbb{CP}^{2^n-1}$ directions
in the state space. A broken unitary labelled by $s$ is a linear combination of 
generalised Pauli matrices with 
weight $w$ with $n-s \leq w \leq n$.

If the penalties of each weight $q_w$ are all different (as in the progressive model),
just the $s=0$  broken unitaries $S^\pm_n$ are penalty eigenstates.
This is the most generic case.
The only exponential horizontal geodesics 
are generated by linear combinations of $S^+_n$ and $S^-_n$.

If some penalties for different weights are degenerate, we can find more eigenstates 
of the penalties which are orthogonal to the unbroken subgroup.
For example, in the draconian model all the weights with $3 \leq w \leq n$
are equally penalised, so all the broken unitaries with $0 \leq s \leq n-3$
generate  projectable exponential geodesics.

 There is a relation between conjugate points in $M$ and $B$   \cite{oneill-geodesics}.
 Let us consider a horizontal geodesic 
 \beq
 \gamma(t): [a,b] \rightarrow M
 \eeq
and let $\gamma(t_0)$ be the first conjugate point of $\gamma$
along the geodesic starting form $\gamma(a)$.
 Then the  projected geodesic $\b(t)=\pi(\gamma(t))$
has a conjugate point for $t_0' \leq t_0$.
 


\section{Towards an exponential complexity}
\label{sect:conjugate points}

The definitions of unitary and state complexity require the minimization of the length of a path connecting the identity with a generic unitary, or the reference state to the target state, respectively.
In the following, we exploit the techniques developed in the previous sections to find explicit classes of geodesics and to find their conjugate points, which play an important role in the minimisation process.

\subsection{Conjugate points and Raychaudhury equation}

An important problem in the geometric approach to complexity 
is to determine the minimal length geodesics that connect the identity to a given unitary.  
From a general result in Riemannian geometry,
a geodesic does not minimise lengths anymore
 after its first conjugate point.
This is not a necessary condition:
there could be a globally shorter path before the first conjugate point.

A useful tool to study conjugate points is the Raychaudhury equation
 (see e.g. \cite{Poisson:2009pwt} for a review).
Let us consider a congruence of geodesics which is orthogonal 
to a family of hypersurfaces in an arbitrary Riemannian manifold. 
Let us denote by $u^\a$ the tangent vector field to the geodesics,
with $u^\a u_\a=1$. The geodesics are in affine parameterization, i.e.
$u^\b D_\b u^\a = 0,$ where $D_{\beta}$ is the covariant derivative.
 The deviation vectors $\xi^\mu $are taken orthogonal to $u^\a$, i.e.
$\xi^\a u_\a=0$.
We can define the transverse part of the metric as:
\beq
h_{\a \b}=g_{\a \b} - u_{\a } u_{\b} \, ,
\eeq
and the tensor
\beq
B_{\a \b}=D_\b u_\a \, ,
\eeq
which can be shown to be symmetric if the congruence of geodesics 
is orthogonal to a family of hypersurfaces.
Morever $B_{\a \b}$ can be decomposed in the trace and traceless part
\beq
B_{\a \b}= \frac{1}{d-1} \Theta \,  h_{\a \b} + \sigma_{\a \b} \, ,
\eeq
where $d$ is the dimension of space,  $\Theta$ is the expansion scalar
and $\sigma_{\alpha \beta}$ the (traceless and symmetric) shear tensor.
The expansion scalar $\Theta$ measures the time derivative of an infinitesimal transverse
volume $\Delta V$ of the geodesic, i.e.
\beq
\Theta= \frac{1}{\Delta V} \frac{d \, \Delta V}{d \lambda} \, .
\eeq
If the scalar $\Theta$ approaches $-\infty$ in some point $r$ along a geodesic,
 it detects the presence of conjugate points
for our congruence of geodesics. This means that the geodesic 
that we are studying does not anymore give us the minimal
distance for points beyond $r$.
The Raychaudhury equation determines the evolution of $\Theta$
along the geodesic flow:
\beq
\frac{d \Theta}{d \lambda} = -\frac{1}{d-1} \Theta^2 -\sigma^{\a \b} \sigma_{\a \b} - R_{\a \b} u^\a u^\b \, ,
\label{raycha-1}
\eeq
where $R_{\a \b}$ is the Ricci tensor and $\l$ is an affine parameter.
There exists also an equation for the traceless part $\sigma_{\a \b}$,
see e.g. \cite{Carroll:2004st}. We discuss this equation in appendix
\ref{shear-appendix}.

\subsection{An application to a simple class of geodesics}

 From eq. (\ref{geogeo}), we know that, in the unitary space,
the exponential of  eigenvectors of the penalty factors matrix $\mathcal{G}$ 
gives us a class of geodesics,
which we call "exponential geodesics".
 It is particularly convenient to apply the
Raychaudhury equation  to this class of geodesics, which have constant $R_{\a \b} u^\a u^\b$.
If we neglect the term $\sigma^{\a \b} \sigma_{\a \b}$ in eq. (\ref{raycha-1}), it can be solved analytically.
 In general this term is non-zero
(see appendix \ref{shear-appendix}), but it is positive definite.
So,  neglecting the $\sigma^{\a \b} \sigma_{\a \b}$ term gives us 
an upper bound for the presence of a conjugate point along a geodesic.

Let us first solve eq. (\ref{raycha-1}) in the limit  $\Theta \rightarrow \infty$, as it is the case for
a family of geodesics starting from the same point. In this case we can 
neglect $R_{\a \b} u^\a u^\b$, leading to:
\beq
\dot{\Theta}+ \frac{1}{d-1} \, \Theta^2 = 0 \, , \qquad \Theta=\frac{d-1}{ \l - k} \, ,
\eeq
where $k$ is an integration constant. This approximation is the same as considering the flat 
space limit. In order to consider a family of geodesics
which start at the same point at $\l=0$, we set $k=0$.
Let us now consider
\beq
\dot{\Theta}+ \frac{1}{d-1} \, \Theta^2 +B =0 \, , 
\eeq
where $B=R_{\a \b} u^\a u^\b$.
The conjugate point, in this approximation, shows up 
only for $B>0$.  Requiring that at small $\l$ the solution reproduces
the flat space one $\Theta=(d-1)/ \l$, we find:
\beq
\Theta=\sqrt{B (d-1)} \cot \left(\sqrt{\frac{B}{d-1}} \,  \lambda \right) \, ,
\eeq
and so it has a conjugate point at 
\beq
\l_0=\frac{\pi \sqrt{d-1}}{\sqrt{B}} \, .
\label{piccola-stima}
\eeq
Since  $\sigma^{\a \b} \sigma_{\a \b}$ is a positive-definite quantity,
the value of $\l_0$
provides  an upper bound for the distance $\l_c$ of the conjugate point from the origin:
\beq
\l_c \leq  \l_0 = \frac{\pi \sqrt{d-1}}{\sqrt{R_{\a \b} u^\a u^b}} \, .
\label{piccola-stima-2}
\eeq
Note that, keeping the Ricci curvature fixed,  $\l_0$
scales exponentially with the number of qubits due to the factor
$ \sqrt{d-1}\approx 2^n$.
This is a first evidence of the exponential 
nature of the maximal complexity.


\subsection{One qubit}
\label{sect-Geodesics_one_qubit}

In order to make the discussion concrete with a clear example
we will consider the one qubit case, see section \ref{1qubit-unitaries}.
In this case the unitary manifold is  a generalised Berger sphere
and an explicit expression for the metric is available.
Introducing the coordinates $(\theta_x,\theta_y,\theta_z)$ to parameterize the unitary
\beq
U=e^{i \s_z \theta_z} e^{i \s_y \theta_y} e^{i \s_x \theta_x} \, ,
\label{UUU}
\eeq
the metric can be written explicitly: 
\beq
g_{ij}=\frac12
\left(
\begin{array}{ccc}
 \Xi & \Psi &  2 P \sin 2 \theta_y \\
\Psi & \Sigma & 0 \\
 2 P \sin 2 \theta_y & 0 & 2 P \\
\end{array}
\right) \, ,
\label{generalised-berger}
\eeq
where
\bea
\Xi &=& 2 \left(P \sin^2 2 \theta_y +\cos^2 2 \theta_y \left( Q \sin^2 2 \theta_z +\cos^2 2 \theta_z \right)\right) \, , \nl
\Psi &=& (1-Q) \cos 2 \theta _y \sin 4 \theta _z \, , \nl
\Sigma&=& (Q-1) \cos 4   \theta _z+Q+1 \, .
\eea
We know from the general analysis that the exponentials of
$\s_x,\s_y,\s_z$ are geodesics, with
\bea
&& G_x: \qquad
\theta_x=\lambda, \, \qquad  \theta_y=\theta_z=0 \, ,
\nl
&& G_y: \qquad
\theta_y=\frac{\lambda}{\sqrt{ Q}}, \, \qquad  \theta_x=\theta_z=0 \, ,
\nl
&& G_z: \qquad
\theta_z=\frac{\lambda}{\sqrt{ P}}, \, \qquad  \theta_x=\theta_y=0 \, ,
\nl
\label{simple-G}
\eea
as can be also checked directly from the geodesic equations of the metric
(\ref{generalised-berger}).

We have seen that the presence of conjugate points on this simple
class of geodesics can be detected  by the Ricci tensor:
 \bea
 R_{x}&=&\frac{2 (1-(P-Q)^2)}{P Q} \, , \nl
  R_{y}&=&\frac{2(Q+P-1)(Q-P+1)}{P  Q} \, ,  \nl
    R_{z} &=& \frac{2(P+Q-1)(P-Q+1)}{P Q} \, ,
\label{1qubit-Ricci}
\eea
where we denote $R_{x,y,z}\equiv R_{\s_x,\s_y,\s_z}$.

\begin{figure*}
\begin{center}
\includegraphics[scale=0.45]{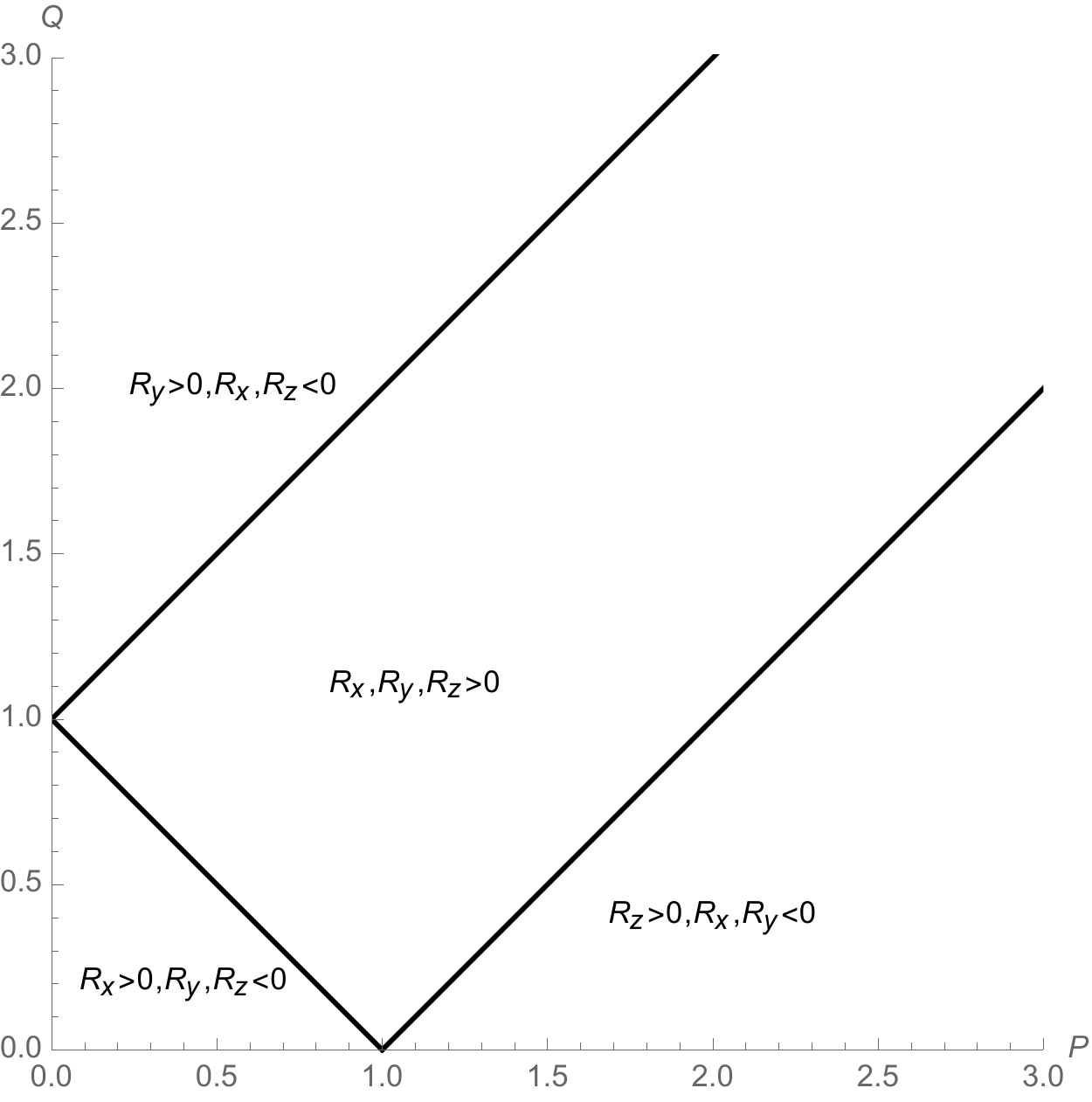}
\qquad \includegraphics[scale=0.45]{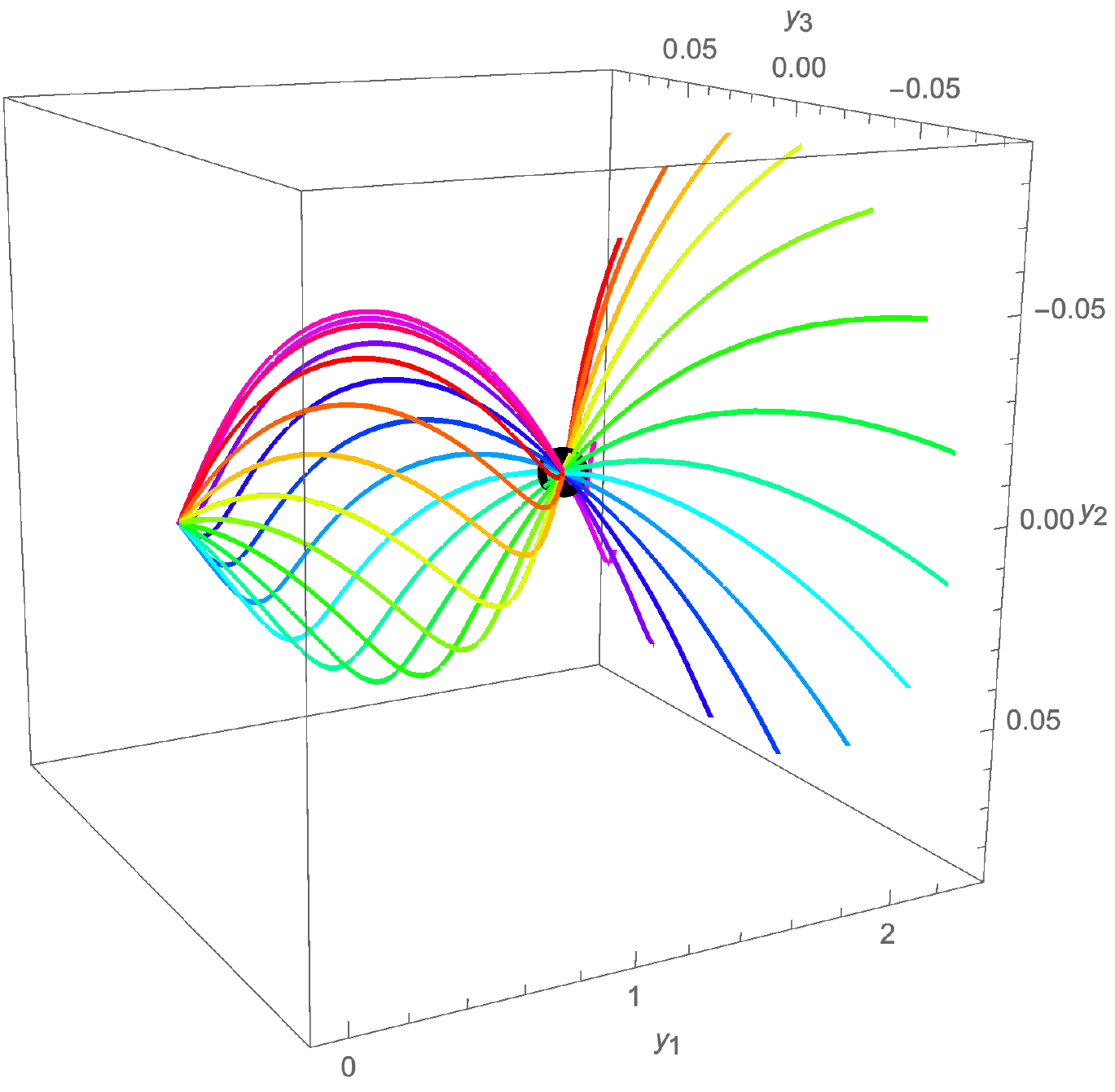} 
\caption{
\label{conjugate-points-fig-1qubit}
Left: Regions where each $R_{x,y,z}$ is positive.
Right: Example of an exact conjugate point (the black spot) of geodesics for $P=Q=0.4$
in stereographic projection. 
}
\end{center}
\end{figure*}

Conjugate points of the geodesic $G_k$ in eq. (\ref{simple-G})
occur in  the regions of the parameter space $(P,Q)$
 where the corresponding $R_k$ is positive,
see Fig. \ref{conjugate-points-fig-1qubit}.
In particular, each of the geodesics $G_k$ for $k=x,y,z$
develops a conjugate point in the region where $R_k > 0$
for
\beq
\l_c \leq \lambda_0 \, , \qquad \lambda_0= \frac{\pi \sqrt{2}}{\sqrt{R_k}} \, .
\eeq 
A plot of an example of conjugate point   is shown 
in Figure \ref{conjugate-points-fig-1qubit} in stereographic projection.

Using eq. (\ref{raycha-2}), it is also possible to include  the $\sigma^{\a \b} \sigma_{\a \b}$ corrections,
in order to determine in general the exact location of the conjugate points.
From such an equation, we can show that $\sigma^{\a \b}$ 
vanishes for $G_x$ in the $P=Q$ case,
for $G_y$ in the $P=1$ case
and for $G_z$ in the  $Q=1$ case (see Appendix \ref{shear-appendix}).  We have then a few 
exact results:
\begin{itemize}
\item For $Q=1$, $G_z$ has a conjugate point at   $\lambda=\frac{\pi}{\sqrt{ P}}$ 
\item For $P=1$, $G_y$ has a conjugate point at   $\lambda=\frac{\pi}{\sqrt{ Q}}$ 
\item For $P=Q$, $G_x$ has a conjugate point at   $\lambda=\pi P $ 
(see the black spot in Fig. \ref{conjugate-points-fig-1qubit}).
\end{itemize}

In particular, it is interesting to consider the limit
in eq. (\ref{1qubit-draconian}), with $P=1$ and $Q \rightarrow \infty$.
In this case the only exponential geodesic 
with a conjugate point is $G_y$.
 In the limit $Q \rightarrow \infty$ the conjugate point moves very close to the origin,
at $\theta_y=\pi/Q$ and at  $\lambda=\pi/\sqrt{ Q}$.
The $G_y$  geodesic is then minimising only very close to the origin,
and the limit is singular. Indeed we already expected
a singularity from the behaviour of curvatures, see eq. (\ref{curvatures-1qubit-draconian}).
Also, sending the penalty $Q$ to infinity does not 
correspond to getting a big complexity in the $\s_y$ direction:
a shortcut with length scaling as $1/\sqrt{Q}$ is for sure available
just after the conjugate point. This is an indication of low maximal complexity
and it is correlated to
a singular limit in the curvature.

It is also interesting to consider the limit
in eq. (\ref{1qubit-moderate}),
where $P=Q \rightarrow \infty$.
The Ricci curvatures are all positive:
\beq
R_x=\frac{2}{P^2} \, , \qquad R_y=R_z=\frac{4}{P} -\frac{2}{P^2} \, .
\eeq
In this case $G_x$ has an exact conjugate point at $\theta_x=\lambda=\pi P$,
while $G_{y,z}$ have conjugate points for $\lambda \lesssim \pi \sqrt{P/2}$,
which correspond to $\theta_y, \theta_z$ of order $1$.
There is no singularity in geodesic, as expected 
from the curvatures in (\ref{curvatures-1qubit-moderate}).
Note that, while the distance of the conjugate point in
$G_{y,z}$ diverges, their position in the coordinate $\theta_{y,z}$
approaches a finite limit for $P \rightarrow \infty$.
The limit of large penalty indeed may correspond
to a large maximal complexity, because no obvious shortcuts are available.
This is supported by numerical computations:
the points with large complexity lay nearby the conjugate point,
and so the maximal complexity scales as $\sqrt{P}$.

In the one qubit case, the exponential geodesics on unitary space
can be projected to the states space using the submersion,
as explained in section \ref{sect:submersions-geodesics}.
Taking as a reference state $|0\rangle$, 
the unbroken subgroup is generated by $\s_z$.
The geodesics shooted  in the orthogonal directions
$\s_x$ and $\s_y$ are then horizontal and projectable.
For generic $P,Q$ there are then two exponential 
horizontal geodesics. The corresponding geodesics on states can be
obtained by the projection of these curves by the submersion $\pi$.

It is more intuitive to plot the geodesics  in the states space, 
since it is a 2-dimensional space.
In the one qubit case, the metric for states in the standard Bloch sphere coordinates  $(\theta,\phi)$ 
is
\beq
g_{ij}=\frac{1}{\Psi} \left(
\begin{array}{cc}
   \Lambda_{11}  &   \Lambda_{12} \\
  \Lambda_{21} &   \Lambda_{22}  \\
\end{array}
\right)
\label{states-metric} 
\eeq
where
\bea
\Lambda_{11} &=&  P \cos ^2 \theta  \cos^2 \phi +P Q \cos ^2 \theta  \sin^2 \phi +Q \sin ^2 \theta  \, ,
\nl
\Lambda_{12} &=&  \Lambda_{21} = P (Q-1) \sin \theta  \cos \theta  \sin \phi  \cos \phi   \, , 
\nl
\Lambda_{22} &=& P \sin ^2 \theta    \left(Q \cos^2 \phi +\sin^2 \phi  \right)  \, , 
\nl
\Psi &=& 4 \left\{ \sin ^2 \theta  \sin^2 \phi +P \cos ^2 \theta +Q \sin ^2 \theta  \cos^2 \phi   \right\}\, . \nl
\eea
We checked  numerically that the projection of the horizontal geodesics in the unitary space
corresponds to geodesics in the states space, as is required by
general results on submersions.

\begin{figure*}
\begin{center}
\includegraphics[scale=0.52]{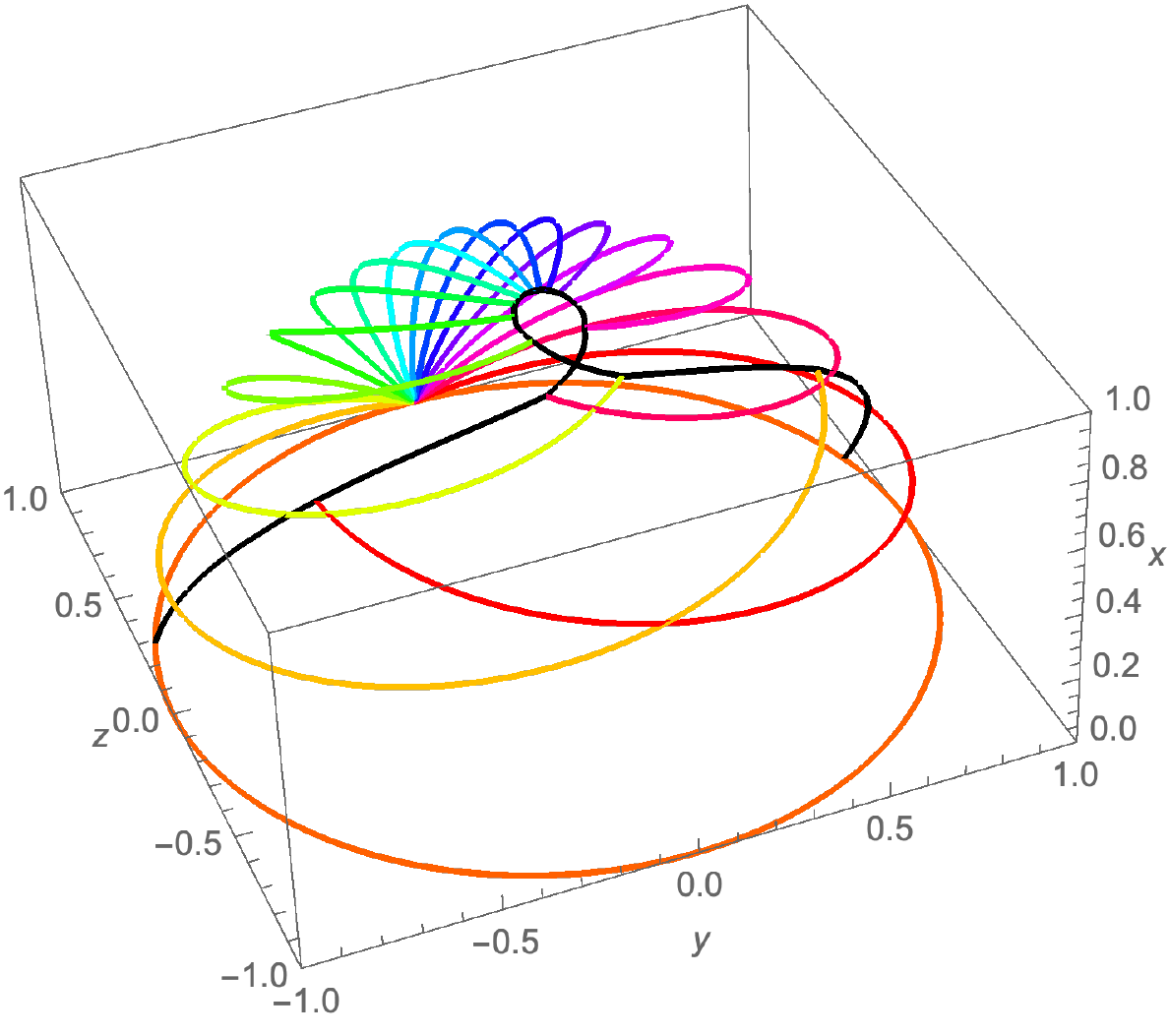} 
\caption{ \label{Q10P10}
Geodesics with length $\l=2.5$ for $Q=10$, $P =10$. The geodesics are plotted in
different colours. The endpoints of the various curves are represented in black. }
\end{center}
\end{figure*}

It is then interesting to plot the geodesics for the case of large $P$ and $Q$
in the state space. In Fig. \ref{Q10P10} the geodesics for the case  $P=Q=10$ 
on the Bloch sphere are shown. In particular, we see that the maximal complexity region lies just
before the conjugate point in $\s_y.$ 
Such a point lies inside the \emph{drop} delimited by the
self intersection of the black curve.
 As it is clear from the figure, no geodesics of length less than $\lambda$ can
penetrate inside  the drop.

\subsection{Draconian model}

 In order to study conjugate points  
 in the draconian model we can use the results from \cite{Nielsen-Dowling} 
 for $R_\s$, where $\s$ is a generalised Pauli matrix with weight $w$:
    \begin{widetext}
\bea
w &=&1 \, , \qquad R_{\sigma } = 2 \le 3 n -2 \ri + \frac{1}{q^2} \le \frac{4^n}{2} - 2 \le 3 n -2 \ri \ri \, , \nl
w &=& 2 \, , \qquad
R_{\sigma } = - 24 \, q \le n-2 \ri + 8 \le 6 n - 11 \ri + \frac{1}{q^2} \le \frac{4^n}{2} - 8 \le 3 n - 5 \ri \ri \, , \nl
w &=& 3 \, , \qquad
R_{\sigma } = \frac{1}{q} \le 12 \, q^2 + \frac{4^n}{2} + 36 \le n -3 \ri - \frac{1}{q} 12 \le 3 n-8 \ri  \ri \, , \nl 
w &\geq& 4 \, , \qquad 
R_{\sigma }  = \frac{1}{q} \le \frac{4^n}{2} + 4 w \le 3 n - 2 w \ri - \frac{1}{q} 4 w \le 3 n - 2 w \ri  \ri\, .
\label{RICCIONE}
\eea
   \end{widetext}
These expressions are valid for arbitrary $n$ and $q$. In particular,
for $q=1$ we recover the cases with uniform penalties $q(w)=1$,
where all the $R_\s$ are the same, i.e. $R_\s=4^n /2$.
In order to have negative scalar curvature, we have 
to scale $q$ with $n$ as $q \approx \mathcal{O}(4^n)$.  

In studying conjugate points along the exponential geodesics,
 it is interesting to consider not only the distance $\l$ from the origin,
 but also their position in a coordinate $\theta$, which runs along the geodesic 
 and does not scale with the penalty. We can define $\theta$ as the length
 in the case with all the penalties $q_\s=1$  (bi-invariant metric). 
In our normalization,  an exponential geodesic can be described by
 \beq
 U(\theta)=\exp \frac{ i \, \theta  \, \s}{\sqrt{2^n}} \, ,
 \eeq
 where $\s$ is the generalised Pauli matrix pointing in the direction
 of the given exponential geodesic.
 Note that this geodesic comes back to the identity matrix after
 a period 
 \beq
 \theta_p=\pi \, 2^{1+n/2} \, .
 \label{theta-p}
 \eeq

 In the large $n$ limit of the unpenalised case $q(w)=1$, 
  eq. (\ref{piccola-stima-2}) gives that 
 for all the weights $w$ 
 \beq
 \l_0=\theta_0 \approx \pi \sqrt{2}  \, .
 \eeq
 In every direction, the cut point must then
 be realised for $\l \leq \l_0$.
 This implies that the maximal 
 complexity is less than $ \pi \sqrt{2} $, which is independent of $n$.

 Let us now consider the regime with negative scalar curvature $q \approx \mathcal{O}(4^n)$.
We can use eq. (\ref{piccola-stima-2}) with $d=4^n-1$
to get an estimate of the distance of conjugate points from the origin:
   \begin{widetext}
\bea
w &=&1 \, , \qquad  \lambda_0 \approx \frac{\pi 2^n }{\sqrt{6 n} } \, ,  \qquad \theta_0=\lambda_0 \, ,\nl
w &= &3 \, , \qquad
 \lambda_0 \approx \frac{\pi 2^n }{\sqrt{12 q} } \approx \frac{\pi}{\sqrt{12}} 
 \, ,  \qquad \theta_0=\frac{\lambda_0}{\sqrt{q}} \approx \frac{1}{2^n}\frac{\pi}{\sqrt{12}} \, , \nl 
w &\geq& 4 \, , \qquad 
 \lambda_0 \approx \pi \sqrt{2 q} \approx \sqrt{2} \pi \,  2^n \, ,
  \qquad \theta_0=\frac{\lambda_0}{\sqrt{q}} \approx \sqrt{2} \pi \, , 
\eea
   \end{widetext}
where  we have inserted $q \approx 4^n$ 
and $\theta_0$ is the length of these geodesics in the unpenalised metric $q_\s=1$.

The geodesics with $w=1$ have a conjugate point
after a length which is exponential in $n$.
However, this cannot correspond to a cut point.
Indeed this conjugate point occurs after that the 
geodesic has passed through the identity matrix many times,
since from eq. (\ref{theta-p}) we have $\theta_p \ll \theta_0$ at large $n$.

The geodesics with $w \geq 4$ instead have 
a conjugate point at $\theta$ of order $1$,
with a length which scales exponentially in $n$.
If in addition we would know that the cut point coincides
with the conjugate point, this would be a proof 
that maximal complexity is scaling 
exponentially with $n$. 
Unfortunately, we don't have a strong indication that this happens. 
Still, the fact that $\theta$ remains of order $1$
makes the possibility that the cut point coincides with conjugate
point not as unrealistic as in the $w=1$ case.
Note that, with very good approximation,  there is no dependence
on $w$ for $w \geq 4$ in the distance of the conjugate point from the origin.

The geodesics with $w=3$ have a conjugate
point at a value of the coordinate $\theta$
very close to the identity. 
In this limit we have evidence that
the conjugate point is also a cut point, because
it happens at infinitesimal  value of the coordinate $\theta$.
However, the distance from the origin is of order $1$,
so this does not teach us anything interesting
 about the possible exponential growth of  complexity at large $n$.
 Also,  the exponential dependence
 $\theta \propto 2^{-n}$ shows that draconian penalties are
 by construction singular.
 
 From the results in section \ref{sect:submersions-geodesics},
 we know that we can find many directions orthogonal to
 the unbroken subgroup which are also penalties eigenvectors.
 In particular, all the tangent directions orthogonal to the vertical space
  with  0$ \leq s \leq n-3$,  see eq. (\ref{s-proiettabile}), 
 contain just operators with weight $w \geq 3$ and so
 generate exponential projectable geodesics.
 The considerations about conjugate points for 
 these exponential geodesics can then be extended 
 to the state space, with the caveat that 
 the conjugate point might occur before in the state space, see  \cite{oneill-geodesics}.

\subsection{Progressive model}

At leading order in $\alpha$, the Ricci contraction
with the unit vector pointing in the direction $\s$ is (for $w>1$)
\beq
 R_{w} = 2 w  \le 2^{w-1} - 2^{2n - 2w +1} \ri \, ,
 \label{wricci}
\eeq
and $R_w=2$ for $w=1$ (see appendix \ref{progressive-appendix}, eq (\ref{ricci-M-appendix})).
This is positive for $w=1$ and for 
\beq
w>\frac{2}{3} (n+1) \, .
\label{large-w}
\eeq

The conjugate point for $w=1$ is estimated at
\beq
\l_0=\theta_0=\frac{\pi}{\sqrt{2}} 2^n \, .
\eeq
Again, eq. (\ref{theta-p}) tells us that
 $\theta_p \ll \theta_0$ at large $n$,
 so this conjugate point cannot 
 correspond to a cut point.
 
 The conjugate points for the generators
 at large $w$ in eq. (\ref{large-w}) are more interesting.
 In this class, the largest positive $R_w$ is at $w=n$,
 which reads
 \beq
 R_n=n (2^n-4) \, ,
 \eeq
and gives a conjugate point at
\beq
\l_0 = \frac{\pi \, 2^{n/2}}{\sqrt{n}} \, , \qquad
 \theta_0=\frac{\l_0}{\a^{n/2}} = \frac{\pi \, 2^{n/2}}{\sqrt{n} \,  \a^{n/2}} \, .
 \label{weight1}
\eeq

The smallest positive value of $R_w$ is realised for 
slightly different values of the integer $w$, depending
on the value of $n$ modulo $3$.
We have to distinguish the following cases:
   \begin{widetext}
\bea
 n=3 a \, , \qquad w&=&2 \frac{n}{3} +1 \, , \qquad
R_w=2^{2 n/3} \le 1+\frac{2 n}{3}\ri \approx n \, 2^{2 n/3} 0.67 \, , \nl 
n=3 a+1 \, , \qquad w&=&\frac{2 n+4}{3}  \, , \qquad
R_w=2^{2 n/3} \le n+2 \ri 2^{1/3} \approx n \, 2^{2 n/3} 1.26 \, , \nl 
n=3 a+2 \, , \qquad w&=&\frac{2 n+5}{3}  \, , \qquad
R_w=2^{2 n/3} \frac{7(2 n+5)}{6 \cdot 2^{1/3}} \approx n \, 2^{2 n/3} 1.85 \, ,
\eea
   \end{widetext}
where $a$ is an integer number.
In all cases $R_w \approx n \, 2^{2 n/3} $ up to order one factors.
This gives a conjugate point at:
\beq
\l_0 = \frac{\pi \, 2^{2 n/3}}{\sqrt{n}} \, , \qquad
 \theta_0=\frac{\l_0}{\a^{n/3}} = \frac{\pi \,  2^{2 n/3 }}{\sqrt{n} \,  \a^{n/3}} \, .
  \label{weight2}
\eeq
Intermediate values of the weight give conjugate points distances which scale
in between the ones in eqs (\ref{weight1}) and (\ref{weight2}).

In order to have small $\theta_0$ 
in the large $n$ limit in eqs. (\ref{weight1}) and (\ref{weight2}), we
have just to require $\a>4$.
The required value of $\a$ should also be large enough
to trust the leading order result (\ref{wricci}).
 The Ricci curvature indeed seems to converge to
the asymptotic value at large $\a$
 quite fast (see Fig. \ref{alpha-fig} for the Ricci scalar).

Since $\theta_0 \rightarrow 0$, we expect that,
for large $n$, the geodesics in eq. (\ref{weight1}) and (\ref{weight2})
are truly minimising ones. 
So we find strong indication that in this limit
the distance of the cut point of the geodesics with large $w$
(in the window $\frac{2}{3} n<w<n$)  is in between 
\beq
\frac{\pi \, 2^{n/2}}{\sqrt{n}}  \leq \l_0 \leq \frac{\pi \, 2^{2 n/3}}{\sqrt{n}} \, .
\eeq
Consequently, the maximal complexity  is bigger than
\beq
 \l_{\rm max} = \frac{\pi \, 2^{2 n/3}}{\sqrt{n}} \, ,
\eeq
and  scales exponentially in $n$.

One may wonder if this is just an artifact of the large $\a$ limit:
indeed in this regime we expect that
the maximal complexity goes to infinity by construction.
In order to clarify this subtle point, let us consider 
higher order corrections  to $R_w$ and to $\l$.
The order $\a^{-1}$ term vanishes for all the Ricci,
except for $w=2$ which is not interesting for conjugate points
(see appendix \ref{progressive-appendix}).
So we need to go to order $\a^{-2}$.

To make the computation simpler, let 
 us consider $w=n$.
In this case, the non-vanishing 
$\a^{-2}$ terms in the sectional curvatures 
which contribute to $R_n$ are:
\bea
&& \delta K(n,2,0) = -\frac{3}{\a^2} \, , \qquad  \delta  K(n,3,0)= \frac{2}{\a^2} \, , \nl
&& \delta  K(n,n-2,0) = -\frac{2}{\a^2} \, , \qquad \delta  K(n,n-1,0)=\frac{1}{\a^2} \, , \nl
&& \delta  K(n,N,1) = \frac{1}{\a^2} \, , \qquad {\rm for} \,\,\,\, 4 \leq N \leq n-1 \, . 
\eea
A direct calculation gives
\bea
R_n&=&
n (2^n-4) +\frac{(n-1) n \left(\left(2^n-16\right) n-2 \left(2^n-4\right)\right)}{6 \alpha ^2} 
\nl
&\approx& n \cdot 2^n +\frac{1}{6 \a^2} n^3 2^n \, .
\eea
The length of the geodesic built from the exponential of a $w=n$ generator
 before the conjugate point  is then, at the next order in $\a$:
 \beq
 \l_0 = \frac{\pi \cdot  2^n}{\sqrt{n \cdot 2^n +\frac{n^3}{6 \a^2} 2^n }} 
 \approx \frac{\pi}{\sqrt{n}} 2^{n/2} \left(  1-\frac{1}{12 \a^2} n^2 \right) \, .
 \eeq
 In order to trust the approximation, we should just 
increase $\a$ in a way slightly faster than $n$ for large $n$, for example quadratically.
From this  polynomial increase of $\a$ with $n$,
we get an exponential increase of complexity.
We believe that this is a strong indication that
maximal complexity scales exponentially with $n$
with a progressive choice of penalties.

This is not a rigorous proof.
 For example, we neglected
the shear term in the Raychaudhuri equation
which may cause the conjugate point to appear before.
 It would be interesting to improve the analysis studying
 the impact of these terms. We leave this as a problem
 for future investigation.

 From section \ref{sect:submersions-geodesics},
 we know that for the progressive model there is just
 a two dimensional space of vectors 
 which are both orthogonal to the vertical space
 and also eigenstates 
 of the penalty. They are generated by arbitrary linear combinations
 of $S^\pm_n$
 in eq. (\ref{Spm}) and they have both $w=n$.
 So the previous calculation in unitary space
 for $w=n$ applies also for state complexity, with the caveat that 
 the conjugate point might occur before in the state space, see  \cite{oneill-geodesics}. 


\section{Conclusions}
\label{sect:conclusions}

In this paper we studied several aspects of complexity geometry.
Using the formalism introduced 
in \cite{Nielsen-Dowling}
for unitary complexity of a  system of $n$ qubits, we showed that
the negativity of sectional curvatures $K$ along the directions $\rho, \s$
in the unitary space is directly related to 
a large penalty factor for the commutator $[\rho, \s]$, i.e.
\bea
&& K \le \rho \, , \sigma \ri = \, \frac{1}{q_{\rho} \, q_{\sigma}} \times 
\\
&&  \left[ -3 \, q_{ \left[ \rho \, , \sigma \right]} + 2 \le q_{\rho} 
 + q_{\sigma} \ri + \frac{\le q_{\rho} - q_{\sigma} \ri^2}{q_{ \left[ \rho \, , \sigma \right]}} \right] \, .
 \nonumber
\eea
In this equation, the only negative term is proportional to the penalty of
$[ \rho \, , \sigma]$, so that in order to get a negative $K(\rho,\s)$
 the penalty $q_{[ \rho \, , \sigma]}$ has to dominate
compared to $q_\rho$ and $q_\s$; this is always possible
for large enough $q_{[ \rho \, , \sigma]}$. 
From this expression it is clear that negative curvature is 
always associated to commutators
of the form
\beq
 \left[ {\rm easy} \, , {\rm easy} \right] = {\rm hard} \, ,
 \label{commutatore-curvatura-negativa}
\eeq
where easy and hard refer to small and large penalty factors respectively.
This is consistent with
the analysis in \cite{Brown:2019whu}. 
The correlation between negative curvature and 
the condition in eq. (\ref{commutatore-curvatura-negativa})
holds  also in the qtrit example
 that we studied in appendix \ref{sect-qudit_3states}.

We applied the formalism of \cite{Nielsen-Dowling} to various examples,
both for small and large number of qubits. 
The one qubit case is already an interesting nutshell for some generic properties
 (see section \ref{1qubit-unitaries}).
First of all, one qubit is a universal closed  subsector of the $n$-qubits space,
because sectional curvatures inside each qubit space depend just on the penalties 
of this subsector. In the generic one qubit parameter space, we have that at least $2$ out of $3$
of the sectional curvatures  in the orthogonal basis are positive.
This argument shows that, for complexity geometry of $n$ qubits,
at least some sectional curvatures are always positive.
Moreover, some of the possible behaviours that 
are realised when some of the penalty factors are sent to infinity
generalise to large number of qubits.
There are two prototypical situations:
\begin{enumerate}
\item if the easy 
generators (which are the ones whose penalties are not sent to infinity)
are enough to construct the generic unitary,
the maximal complexity does not diverge.
 Some of the sectional curvatures instead diverge
and the geometry is singular. An example of 
this case is realised for $Q \rightarrow \infty$ and $P$ constant. 
\item if the remaining easy 
generators are {\it not } enough to construct the generic unitary,
the maximal complexity is infinity by construction and the sectional curvatures
do not diverge. An example of this case is  for $P=\b Q \rightarrow \infty$,
with $\b$ constant, where both vanishing (for $\b=1$) and negative
(for $\b \neq 1$) scalar curvatures can be realised. 
\end{enumerate}

For a larger number of qubits $n$ the situation is much more intricate,
because the dimension of the space of unitaries scales as $4^n$.
The allowed values of sectional curvatures in the orthogonal basis
have large multiplicities, which can scale exponentially or polynomially with $n$
and the weight $w$.
In appendix \ref{counting-appendix} we provide general expressions
for this counting.
For large $n$ we have a huge arbitrariness in the choice of the penalty factors.
Two useful  prototypes are:
 \begin{itemize}
\item draconian  penalties, defined by eq. (\ref{draconian-q-intro}).
In the large $q$ limit, for fixed $n$,  complexity does not diverge,
and the geometry becomes singular. This is similar to point 1 of the one qubit case.
\item progressive penalties, as defined in eq. (\ref{progressive-intro}).
In the large $\a$ limit,  complexity diverges for fixed $n$
and the geometry is not singular (the sectional curvatures scale as $\a^0$).
The scalar curvature, see eq. (\ref{scalar-curvature-progressive}), is negative.
The situation is similar to point 2 in the one qubit case. 
\end{itemize}

So far we discussed  complexity as  defined for unitary operators.
For applications to holography, it is more relevant to 
consider the different but somehow related notion of state complexity  \cite{Susskind:2014moa}.
Complexity for states is defined as the lowest
possible complexity of an operator which prepares
the state, starting from a given reference state. In general,
we have to minimise over all the possible unitaries
that prepare the given state \cite{Brown:2019whu}.
The complexity metric here is much more intricate, because
the geometry is not homogeneous. 

In section \ref{sect:states-submersion}, we point out that the 
relation between the unitary and the state geometry
follows directly from the mathematical theory of Riemannian 
submersions \cite{oneill,oneill-geodesics,besse}.
In particular, the geodesics in the state space $B$
can be found by a projection of a particular
class of geodesics (the horizontal ones)
from the unitary space $M$. Moreover, conjugate points for geodesics
in $B$ are realised for a complexity equal or less than the
one in $M$. Curvatures in the state and in the unitary spaces
are related by O'Neill's formula \cite{oneill}.
Geodesics in the state space 
can be in principle computed without even knowing 
the metric on $B$.
Our approach gives also a a closed-form expression for the state metric.
We checked that this result reproduces the known
 1-qubit metric with arbitrary penalties. 
As a new  application, we determine the state complexity metric
and curvatures for the one qutrit example.

An important open problem is to understand the regime
in which the complexity metric provides a complexity distance
scaling exponentially with the number of qubits.
In section \ref{sect:conjugate points} we provide robust evidence
for the exponential behaviour of complexity
for  progressive penalties.
The analysis is based on the study of conjugate points in the unitary space.
For a general manifold,  the study of conjugate points 
does not provide direct information about the maximal possible complexity,
because a geodesic might cross its cut point before the conjugate point.
This obstruction can be circumvented if one considers
parametric regimes in which the angular position of the 
conjugate point approaches the identity.
In this limit we expect that the cut point coincides with
the conjugate point. We show that this regime 
is realised for progressive penalties at large $\a$
and we give an estimate for a lower bound
for the scaling of complexity.
This bound scales exponentially with $n$.


 \section*{Acknowledgments}
 
 We are grateful to Mauro Spera for very precious geometrical insights.
 We thank Luca Cassia, Alice Gatti and Alessandro Tomasiello
  for valuable discussions.
 
 S.B. acknowledges support from the Independent Research Fund Denmark 
 grant number DFF-6108-00340 "Towards a deeper understanding of black holes with non-relativistic holography".
 A.L. is supported by UKRI Science and Technology Facilities Council (STFC) Consolidated Grants ST/P00055X/1 and ST/T000813/1.
  The work of G.B.D.L. is supported in part by the Simons Foundation Origins of  the  Universe  Initiative  
  (modern  inflationary  cosmology  collaboration)  and  by  a Simons Investigator award.
  N.Z. acknowledges the Ermenegildo Zegna's Group for the financial support.
  

\section*{Appendix}
\addtocontents{toc}{\protect\setcounter{tocdepth}{1}}
\appendix

\section{Counting  non-vanishing sectional curvatures}
\label{counting-appendix}

   Given two generators $(\rho, \sigma)$,
   we define $l$ as the number of corresponding tensorial product entries in
    which $\rho$ and $\sigma$ have different Pauli matrices
     (for anticommuting $\rho$ and $\sigma$, $l$ is odd).
    We define $m$ as the number of corresponding tensorial product 
    entries in which $\rho$ and $\sigma$ have the same Pauli matrices.
    
   The number of entries in the tensorial product
    in which there is a Pauli matrix in    $\sigma$ 
    and an identity in the corresponding entry in $\rho$ is given by
    \beq
    s=N-l-m \, .
    \label{def_s}
    \eeq

Due to the properties of generalised Pauli matrices, if a pair of generators in the basis do not commute, 
then they necessarily need to anti-commute.
 Consequently, the commutator $[\rho,\sigma]$ has weight
 \beq
 w=M+N-l-2 m \, ,
 \eeq
 where $l+m \leq \min (M,N)$.  The minimal weight is realised just for $l=1$ and for $m=\min(N-1,M-1)$.
 The maximum weight instead is realised by $l=1$ and $m=0$.
 
 In order to parameterize the possible values of the weight $w$, let us introduce 
 an integer  label $r$:
 \bea
 &&{\rm for} \,\,\,\, N \leq M \, :  \\
 &&  r=N- \frac{l+1}{2} - m \, , \qquad
    r = 0, \dots, N-1\, ,  \nonumber
 \eea
 \bea
 && {\rm for} \,\,\,\, N > M  \, : \\
   &&   r=M- \frac{l+1}{2} - m \, ,  \qquad r = 0, \dots, M-1\, ,  \nonumber
 \eea
 in such a way that the weight of the commutator is
\beq
w_r= | M-N | +1 + 2 r \, . 
\label{wr}
\eeq
The $r=0$ case corresponds to the lowest possible weight of the commutator,
while the maximum of  $r$  corresponds to the maximum weight.

The weight is limited also by the number of qubits, i.e.
$w_r \leq n$. So, for any given pair $(M,N)$, we must have that the integer $r$
is in the following range:
 \bea
 \label{range-r}
 &&{\rm for} \,\,\,\, N \leq M \, , \\
  && 0 \leq r \leq \min \le N-1, \frac{n-|M-N|-1}{2} \ri  \, , 
   \nonumber
 \eea
 \bea
 && {\rm for} \,\,\,\, N > M  \, , \\
  &&  0 \leq r \leq \min \le M-1, \frac{n-|M-N|-1}{2} \ri \, .
  \nonumber
 \eea
 Note that for each fixed number of qubits $n$, $r \leq \left[ (n-1)/2 \right]$ 
where $[ \dots ]$ denotes the integer part.

Given a generator $\rho$ in the basis with weight $M$,
we similarly denote by $\mathcal{R}(M,N,r)$ the number of generators
with weight $N$ whose commutator with $\rho$ has a weight parameterized 
by a given integer $r$, as in eq. (\ref{wr}).

We give now an explicit formula for $\mathcal{R}(M,N,r)$.
Let us first consider the $N \leq M$ case and let us start with $r=0$.
In this case
we need to determine how many $\s$ will give a $[\rho,\s]$ with the minimal possible weight.
As stressed before, this is realised just for $l=1$, $m=N-1$ and $s=0$. 
We have $M$ places to stick the $l=1$ entry of $\s$ (which corresponds to a different 
Pauli matrix compared to $\rho$, so there is an extra factor of $2$), and then 
we have $\binom{M-1}{N-1}$ ways to stick the $m=N-1$ entries of
$\s$ with the same Pauli matrix as in $\rho$.
The number of such matrices is:
\beq
\mathcal{R}(M,N,0) =  2 M  \binom{M-1}{N-1} \, .
\label{rzero}
 \eeq
 Let us consider  $r=1$. Here in general we have two possible situations.
 We may have $l=1$, $m=N-2$, $s=1$ or instead
 $l=3$, $m=N-3$ and $s=0$. In the first case,
 there are $3$ ways to choose the Pauli matrix in $\s$
 which has an identity in the corresponding entry in $\rho$.
 This gives:
 \bea
 \mathcal{R}(M,N,1) &=&  
 \binom{M}{3} 2^3 \binom{M-3}{N-3}   \\
&+& \binom{M}{1} 2^1  \binom{M-1}{N-2}  \binom{n-M}{1} 3  \, .
\nonumber
 \eea
In the general case we have to sum over all the possible odd values of $l$;
it is then convenient to set $l=2k+1$
with  integer $k$.
 In general we have $ \binom{M}{l}  2^{l}$ ways to set the entries in tensor product
 where $\rho$ and $\sigma$ have different Pauli matrices,
 $ \binom{M-l}{m}$ ways to set the entries in such a way that 
 $\rho$ and $\sigma$ have the same Pauli matrices 
 in the corresponding entries  and
 $ \binom{n-M}{s} 3^{s}$ ways to set entries 
 in which in the corresponding elements of $\rho$ and $\sigma$
there are  an identity matrix and a Pauli matrix respectively.
The total combinatorial factors is
 \bea
&&\mathcal{R}(M,N,r) =
\sum_{k=0}^r
 \binom{M}{l}  2^{l} \binom{M-l}{m} \binom{n-M}{s} 3^{s}
\nl
&=&  \sum_{k=0}^r
 \binom{M}{2 k +1}  2^{2 k +1} \binom{M-2 k -1}{ N-k-1-r} \times \nl
&&  \binom{n-M}{r-k} 3^{r-k} \, , \nl
 \label{molteplicita1}
 \eea
where we used $s=r-k$.
In this expression we should not worry about
negative values of $N-k-1-r$, which indeed may occur,
because the corresponding terms in the sum vanish after analitically continuing the binomial coefficients with the $\Gamma$ function.

If $N > M$, we can write a similar formula.
We can still use the same eq. (\ref{molteplicita1}),
with $s=r-k+N-M$ and $m$ accordingly given by (\ref{def_s}):
 \bea
&&\mathcal{R}(M,N,r) =
 \sum_{k=0}^r
 \binom{M}{l}  2^{l} \binom{M-l}{m} \binom{n-M}{s} 3^{s}
\nl
&=&\sum_{k=0}^r
 \binom{M}{2 k +1}  2^{2 k +1} \binom{M-2 k -1}{ M-k-1-r} \times \nl
&&  \binom{n-M}{r-k+N-M} 3^{r-k+N-M}  \, .
 \nl
 \eea
Let us denote by
$\mathcal{N}(M,N,r)$ the number of sectional
curvatures with value given by eq. (\ref{values-K}).
These can be found as
\beq
\mathcal{N}(M,N,r)=\mathcal{N}_M  \mathcal{R}(M,N,r) =
\mathcal{N}_N  \mathcal{R}(N,M,r) \, ,
\eeq
where $\mathcal{N}_M, \mathcal{N}_N$ are defined in eq. \eqref{eq:N_subscript}.


\section{Explicit calculations for the progressive penalties case}
\label{progressive-appendix}

In this section we consider the choice in eq. (\ref{progressive}).
A direct calculation gives, for  $N \leq M$:
\bea
K(M,N,r)=-3 \a^{2(r+1-N)} + 2 \a^{1-N} +2 \a^{1-M} \nl 
+\a^{-2 r} (1+ \a^{-2 (M-N)} - 2 \a^{-(M-N)}) \, , \nl
\label{K-alpha}
\eea
and, for   $N > M$:
\bea
K(M,N,r)=
- 3 \alpha^{2 \le r+1 -M \ri} + 2 \alpha^{1-N} + 2 \alpha^{1-M} \nl
 + \alpha^{- 2 r}  ( 1+ \alpha^{-2 \le N-M \ri} - 2 \alpha^{- \le N-M \ri} ) \, . \nl
\eea
Note that at large $\a$ sectional curvatures
scale at most as $\a^0+\mathcal{O}(\a^{-1})$.

\subsection{Leading order}
Let us start with the $\a^0$ terms. 
For $r=0$, the only non-vanishing sectional curvatures at this order
 are $K=1$, for $M=N=1$ and
\beq
M, N > 1 \, , \qquad M \neq N \, .
\eeq
For $r \geq 1$, the only term that can be of order $\alpha^0$ is for $M, N >1$ and is given by
\bea
&& K \le M , N , r \ri =  \nl
&& =\begin{sistema}
- 3 \alpha^{2 \le r+1 -N \ri} + O \le \alpha^{-1} \ri \qquad M \geq N \\
- 3 \alpha^{2 \le r+1 -M \ri} + O \le \alpha^{-1} \ri \qquad M < N
\end{sistema} \, . \nl
\eea
If $M \geq N$, we have $r = 0, ..., N-1$, then $K = -3$ only for the maximal value $r = N-1$. 
If $M < N$, we have $r = 0, ..., M-1$, then $K = -3$ only for the maximal value $r = M-1$.

We first compute the Ricci 
tensor contracted with a unit vector  $u \le \s \ri$, where $\sigma$ has weight $M$,
as defined in eq. (\ref{ricci-diagonal-def}):
\beq
R_M = \sum_N \sum_r K \le M , N , r \ri \, \mathcal{R} \le M, N, r \ri \, .
\eeq
For $M=1$, the only leading-order
 contribution is for $M=N=1$:
\beq
R_1 =  \mathcal{R} \le 1, 1, 0 \ri= 2 \, .
\eeq
Let us now consider $1 < M \leq n$. 
The positive leading-order contributions to $R_M$
 are given by the scalar curvatures with $r=0$, whose value is $K=1$:
\bea
\label{ricci positive}
R_M^+  &=& \sum_{N=2}^{M-1}  \mathcal{R} \le M, N, 0 \ri  
+ \sum_{N=M+1}^{n}  \mathcal{R} \le M, N, 0 \ri  \nl
 &=& 2 M \left( 2^{M-1} -3 + 2^{2 \le n-M \ri} \right) \, . 
\eea
The negative leading-order contributions to $R_M$ are given by the scalar
 curvatures with $r=N-1$ if $M \geq N$ and $r=M-1$ if $M<N$, all equal to $K= -3$.
 The expression turns out to be the same for both the cases:
 \bea
&& {\rm for} \,\,\, M \geq N  \\
 && \mathcal{R} \le M, N, N-1 \ri  = 2 M \binom{n-M}{N-1} \, 3^{N-1} \, ,
 \nonumber
 \eea
  \bea
&& {\rm for} \,\,\, M < N  \\
 && \mathcal{R} \le M, N, M-1 \ri  = 2 M \binom{n-M}{N-1} \, 3^{N-1} \, , 
 \nonumber
 \eea
We finally get
\bea
\label{ricci negative}
R_M^-  &=& \sum_{N=2}^{1+n-M}  2 M \binom{n-M}{N-1} \, 3^{N-1}  \nl
 &=& -6 M \left[ 2^{2 \le n-M \ri} -1 \right] \, .
\eea
The maximum value of $N$ in the sum, $N_{\rm max} = 1+n-M$, ensures that $r=M-1$ is allowed in the case $M<N$,
as can be obtained from eq. (\ref{range-r}).

The final result for $R_M$ at the leading order is:
\beq
R_M = R_M^+ + R_M^- = 2 M \left(2^{M-1} - 2^{2 \le n-M \ri +1} \right) \, .
\label{ricci-M-appendix}
\eeq
Using eq. \eqref{eq:N_subscript} and this result, the scalar curvature is computed as
\beq
R = \sum_{M = 1}^n \mathcal{N}_M \, R_M= 3 n \le 4^n - 2 \, 7^{n-1} \ri \, .
\label{scalar-curvature-progressive-appe}
\eeq

\subsection{Next to leading order}

We can systematically improve this calculation order by order in 
the expansion parameter $\a$.
For example, at order $\a^{-1}$, the only non-zero contribution to the sectional curvatures, that
we denote as $\delta K(M,N,r),$ are 
\bea
M &=& N=2  \, , \qquad r= 0,1 \, , \qquad \delta K=\frac{4}{\a} \, , 
\nl
M &=& N+1 \, , \qquad  N \geq 3  \, , \qquad r=0 \, , \qquad  \delta K=-\frac{2}{\a} \, ,
\nl
M &=& 2 \, , \qquad N \geq 4 \, , \qquad r=0 \, , \qquad   \delta K=\frac{2}{\a} \, ,
\nl
M &=& 2 \, , \qquad N \geq 3 \, , \qquad r=1 \, , \qquad  \delta K=\frac{2}{\a} \, , 
\nonumber
\eea
and the ones obtained exchanging $M$ with $N$.
Due to a non-trivial cancellation, 
the only corrections to $R_M$  is for $w=2$
\beq
\delta R_2= \frac{4^n}{\a} \, .
 \eeq
 This gives the following correction to the curvature
 \beq
 \delta R=\frac{9}{2} n (n-1) \frac{4^n}{\a} \, .
 \eeq

\section{State complexity for 1 qutrit}

\label{sect-qudit_3states}

In this section we show an application of the method in section \ref{submersion-explicit}
to determine the metric and the curvature 
properties in the space of states, using the explicit decomposition of the unitary space as a submersion.
We consider the case of a qudit theory, which describes a system with $n$ energy levels.
In particular, we focus on the case of
 one qutrit, where $n=3$ and the group manifold is $M=\mathrm{SU}(3).$

The corresponding space of states is $M/G=\mathbb{CP}^2,$ which is parameterized by two complex coordinates $(z_i, \bar{z}_i)$ with $i \in \lbrace 1,2 \rbrace .$ 
Alternatively, we can use four real coordinates $(\theta_i, \phi_i)$ where $\theta_i \in [0,\pi]$ and $\phi_i \in [0, 2 \pi]$ with $i \in \lbrace 1,2 \rbrace . $ 
The parameterization with complex coordinates is useful to transform the reference state, which we conventionally take to be $| \psi_0 \rangle = (1,0,0),$ into the generic state
\begin{equation}
\ket{\psi} = \frac{1}{\sqrt{1+z_i \overline{z}^i}} \begin{pmatrix}
1 \\ z_1 \\ z_2
\end{pmatrix}  =
\begin{pmatrix}
\cos \theta_1 \\
e^{i \phi_1} \sin \theta_1 \cos \theta_2 \\
e^{i \phi_2} \sin \theta_1 \sin \theta_2
\end{pmatrix} \, .
\end{equation}
The parameterization with angular coordinates, which we use in the second equality, will be convenient to describe the curvatures, giving a compact expression for the Ricci scalar.

Here and in the following, the subscript
 refers to the coordinate dependence of the group element from the space of states ($\theta$ subscript)
  or from the additional coordinates that bring to the space of unitaries ($\lambda$ subscript).
Instead the superscript $(K)$ refers to the group $\tmop{SU}(K)$ to which the element belongs.
The generic element of the coset space $M/G$ is given by
   \begin{widetext}
\beq
\begin{aligned}
U_{\theta}^{(3)} & = \frac{1}{\sqrt{1+z_i \overline{z}^i}} 
 \begin{pmatrix}
1  & -\overline{z}_j \\ 
z_i & \sqrt{1+z_i \overline{z}^i} \delta_{ij} -  \frac{z_i \overline{z}_j}{1+\sqrt{1+z_i \overline{z}^i}} 
\end{pmatrix} = \\
& = \begin{pmatrix}
\cos \theta_1 &  -e^{-i \phi_1} \sin \theta_1 \cos \theta_2 & -e^{-i \phi_2} \sin \theta_1 \sin \theta_2 \\
-e^{i \phi_1} \sin \theta_1 \cos \theta_2 & \cos \left( \frac{\theta_1^2}{2} \right) - \cos (2 \theta_2)  \sin \left( \frac{\theta_1^2}{2} \right) & - e^{i(\phi_1-\phi_2)} \sin \left( \frac{\theta_1^2}{2} \right) \sin (2 \theta_2) \\
 -e^{i \phi_2} \sin \theta_1 \sin \theta_2 & - e^{-i(\phi_1-\phi_2)} \sin \left( \frac{\theta_1^2}{2} \right) \sin (2 \theta_2) & 
  \cos \left( \frac{\theta_1^2}{2} \right) + \cos (2 \theta_2)  \sin \left( \frac{\theta_1^2}{2} \right) 
\end{pmatrix} \, .
\end{aligned}
\label{eq:coset_element_SU3}
\eeq
   \end{widetext}
While the last equality is specific of this case, the expression in the first line applies to the space $\mathbb{CP}^{K}$ with $K \in \mathbb{N}$ arbitrary.
In the general case, the only difference is that the index runs over $i \in \lbrace 1, \dots, K \rbrace .$
 
The group $\tmop{SU}(3)$ contains as maximal subgroup $\tmop{SU}(2) \times \tmop{U}(1).$ 
In order to build the stabilizer of the element $(1,0,0)$ inside $\tmop{SU}(3),$ we use a recursive procedure.
The $\tmop{SU}(2)$ factor corresponds to the case of a single qubit: then the stabilizer of the element $(1,0)$ is given by the exponental of the Pauli matrix $\sigma_z,$ which reads
\beq
V^{(2)}_{\lambda} = e^{i \lambda_2 \sigma_z} = \begin{pmatrix}
e^{i \lambda_2} & 0 \\
0 & e^{-i \lambda_2}
\end{pmatrix} \, .
\label{eq:stabilizer_1qubit}
\eeq  
Now we consider the coset element of $\tmop{SU}(2),$ that can be easily taken from the lower-dimensional generalization of eq.~\eqref{eq:coset_element_SU3} and reads
\beq
U_{\lambda}^{(2)} = 
\begin{pmatrix}
\cos \lambda_1 & - e^{-i \lambda_3} \sin \lambda_1  \\
e^{i \lambda_3} \sin \lambda_1 & \cos \lambda_1
\end{pmatrix} \, .
\eeq
In this way we build the generic element of $\tmop{SU}(2)$ as
\bea
U^{(2)} &=& U^{(2)}_{\lambda} V^{(2)}_{\lambda} \nl
&=&  \begin{pmatrix}
e^{i \lambda_2} \cos \lambda_1 & -  e^{-i (\lambda_2+\lambda_3)} \sin \lambda_1 \\
 e^{i (\lambda_2+ \lambda_3)} \sin \lambda_1 & e^{-i \lambda_2} \cos \lambda_1
\end{pmatrix} \, .
\nonumber
\eea
Finally, the stabilizer of the reference state inside $\tmop{SU}(3)$ requires another $\tmop{U}(1)$ factor, coming from a global phase that does not change the physics of the system.
Indeed, we have the freedom to add another real variable, and the generic element of the maximal subgroup can be written as
\beq
V^{(3)}_{\lambda} = p_2 U^{(2)}_E  \, ,
\eeq
with the phasis given by the matrix
\beq
p_{K} = \begin{pmatrix}
e^{i K \lambda_{2K}} & 0  & \dots & 0 \\
0 & e^{- i \lambda_{2K}} & \dots & \dots \\ 
\dots & \dots & \dots & 0 \\
0 & \dots & 0 & e^{-i \lambda_{2K}}
\end{pmatrix} \, ,
\label{eq:phasis_generic} 
\eeq
and where we need to embed the matrix $U^{(2)} $ inside $\tmop{SU}(3)$ as follows:
\beq
U^{(2)}_E \equiv \begin{pmatrix}
1 & 0 \\
0 & U^{(2)}
\end{pmatrix} \, .
\eeq
In this way we finally obtain the stabilizer of the reference state as
   \begin{widetext}
\beq
V_{\lambda}^{(3)} = 
\begin{pmatrix}
e^{2 i \lambda_4} & 0 & 0 \\
0 & e^{i(\lambda_2-\lambda_4)} \cos \lambda_1  & - e^{-i(\lambda_2+\lambda_3+\lambda_4)} \sin \lambda_1  \\
0 & e^{i(\lambda_2+\lambda_3-\lambda_4)} \sin \lambda_1  &  e^{-i(\lambda_2+\lambda_4)}  \cos \lambda_1
\end{pmatrix} \, .
\label{eq:stabilizer_su3}
\eeq
   \end{widetext}
It depends on four real coordinates $\lambda_i,$ with $i \in \lbrace 1,2,3,4 \rbrace .$

Now we want to apply eq.~\eqref{metric-states-general} to determine the metric on the states space starting from the right-invariant form $u_s$ and the left-invariant form $v_s$ defined in \eqref{eq:right_left_inv_forms}.
In addition, we need to specify the penalty matrix $M.$
The most relevant case corresponds to penalizing
 the unbroken generators, because it is a configuration
  that allows for the existence of commutators of the form 
\beq
[\mathrm{easy}, \mathrm{easy}] = \mathrm{hard} \, ,
\eeq
which are expected to generate negative curvature.
This happens due to the algebraic relations \eqref{e-h-2}, which occur because we selected a maximal subalgebra.
In addition, by considering $0\leq P <1,$ we can also realize a relation of the form \eqref{e-h-1},
where only the broken generators are penalized.

For these reasons, we take the penalty matrix to be
\beq
M = \mathrm{diag} (P,P,P,P,1,1,1,1) \, ,
\label{eq:penalty_matrix_SU3}
\eeq
where the first four components refer to directions along the maximal subgroup 
$\mathrm{SU}(2) \times \mathrm{U}(1),$ and the last four directions to the 
broken generators.

We analitically compute the metric on states \eqref{metric-states-general}.
The result is
\bea
&& ds^2_S  =  d \theta_1^2 +
\frac{2P \sin^2 \theta_1}{A(\theta_1)} d \theta^2_2 
+ \frac{2P \sin^2 \theta_1 \cos^2 \theta_2}{A(\theta_1)} d \phi_1^2  \nl
&&+  \frac{C(\theta_1, \theta_2)}{A(\theta_1) B(\theta_1)} d \phi_2^2 
 + \frac{2P \sin^2 \theta_1 \cos^2 \theta_2}{A(\theta_1) B(\theta_1)} D(\theta_1) \times \nl
&& \left( \cos^2 \theta_2 (d \phi_1 -d \phi_2)^2 + 2 d\phi_1 d\phi_2 \right) \, ,
\eea
where we defined for convenience the quantities
\bea
 A(\theta_1) &\equiv& (P-1) \cos (2 \theta_1) + P +1 \, , \nl
B(\theta_1) &\equiv& (P-1) \cos (4 \theta_1) + P +1 \, ,  \nl
 C(\theta_1, \theta_2) &\equiv&  P \sin^2 \theta_1  \left[ B(\theta_1) - \cos (2 \theta_2) \right. \nl
&& \left.  \left( 2P \cos (2 \theta_1) + (P-1) \sin^2 (2 \theta_1) \right) \right] \, ,  \nl
 D(\theta_1) &\equiv&  3(P-1) \cos (2 \theta_1) + P +3 \, . 
\eea
The metric depends on the angles $\theta_i$ but not on the phases $\phi_i.$
The scalar curvature reads
\bea
R &=& \frac{15}{2} \left( \frac{1}{P} -1 \right) + \frac{14 P}{\left( (P-1) \cos^2 \theta_1 +1  \right)^2} \nl 
&+& \frac{2-2P(3P+14)}{(P+3)(P-1)\cos^2 \theta_1 + P +3}  \nl  
&+& \frac{96 P}{\left( (P-1) \cos (4 \theta_1) + P +1 \right)^2} \nl 
&+& \frac{-8(P-1)(9P+19) \cos^2 \theta_1 + 3P (P-18) +3}{(P+3) \left( (P-1) \cos (4 \theta_1) + P +1 \right)} \, . \nl
\label{eq:Ricci_scalar_CP2}
\eea
We observe that the Ricci scalar depends only on the angular coordinate $\theta_1,$ 
giving a further simplification with respect to the metric on $\mathbb{CP}^2.$
This is due to the many symmetries of the 
  penalties in eq. (\ref{eq:penalty_matrix_SU3}).

\begin{figure*}[ht]
\centering
\includegraphics[scale=0.3]{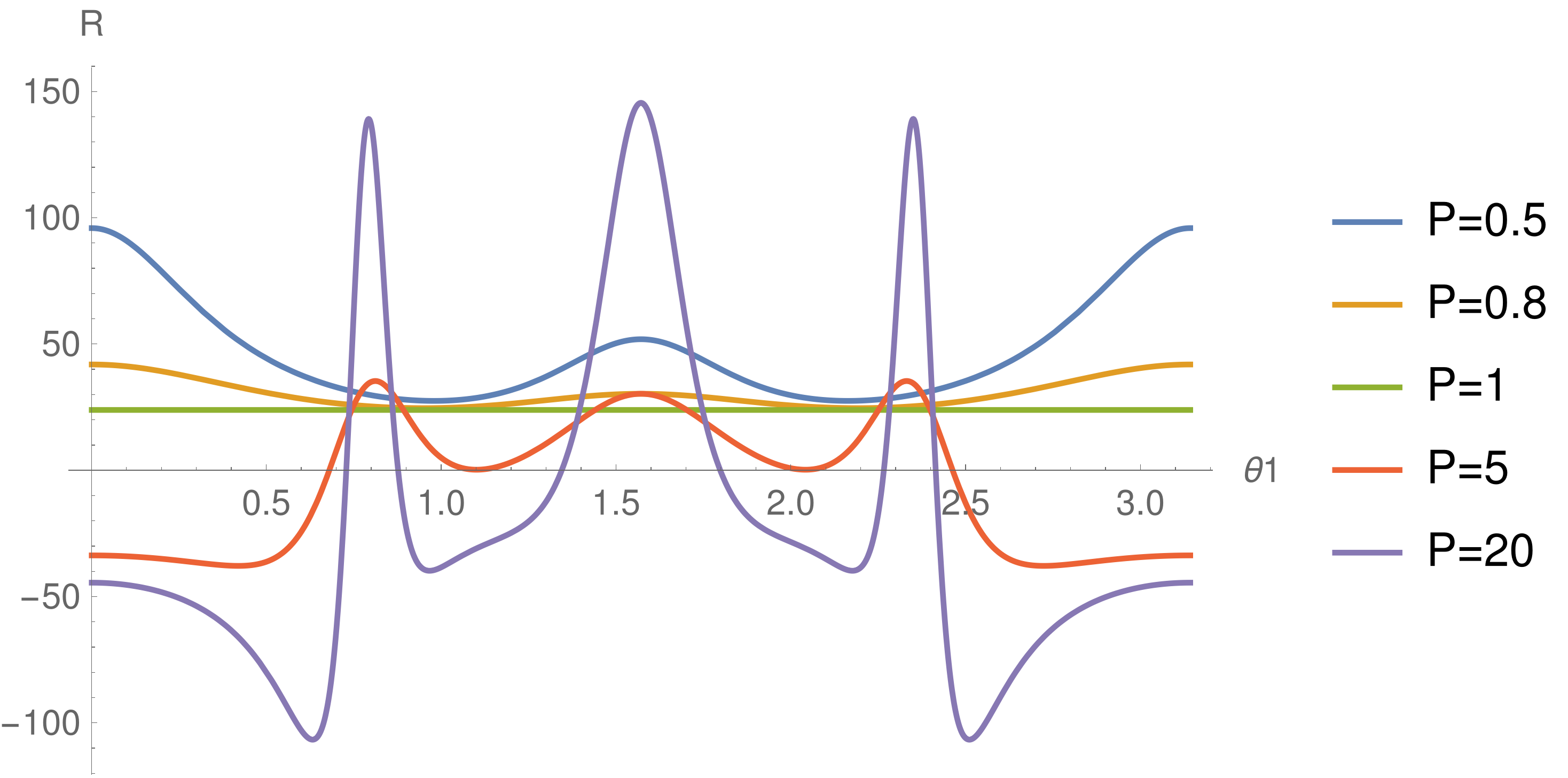}
\caption{ \label{fig:R_CP2}
Scalar curvature \eqref{eq:Ricci_scalar_CP2} for the state space 
$\mathbb{CP}^2$ with penalty factors $P$ applied to all the generators of the maximal subgroup.}
\end{figure*}

In Fig. \ref{fig:R_CP2} we plot the Ricci scalar as a function of $\theta_1$
 for different values of the penalty $P.$
We observe that when $0 < P <1$ the scalar curvature is always positive, and reaches a constant value $R=24$ when
$P=1,$ the case of undeformed inner product on $\mathrm{SU}(3).$
When $P>1$ there is always a region with negative curvature which increases its size accordingly to the increasing of the penalty.

\begin{figure*}[ht]
\centering
\includegraphics[scale=0.25]{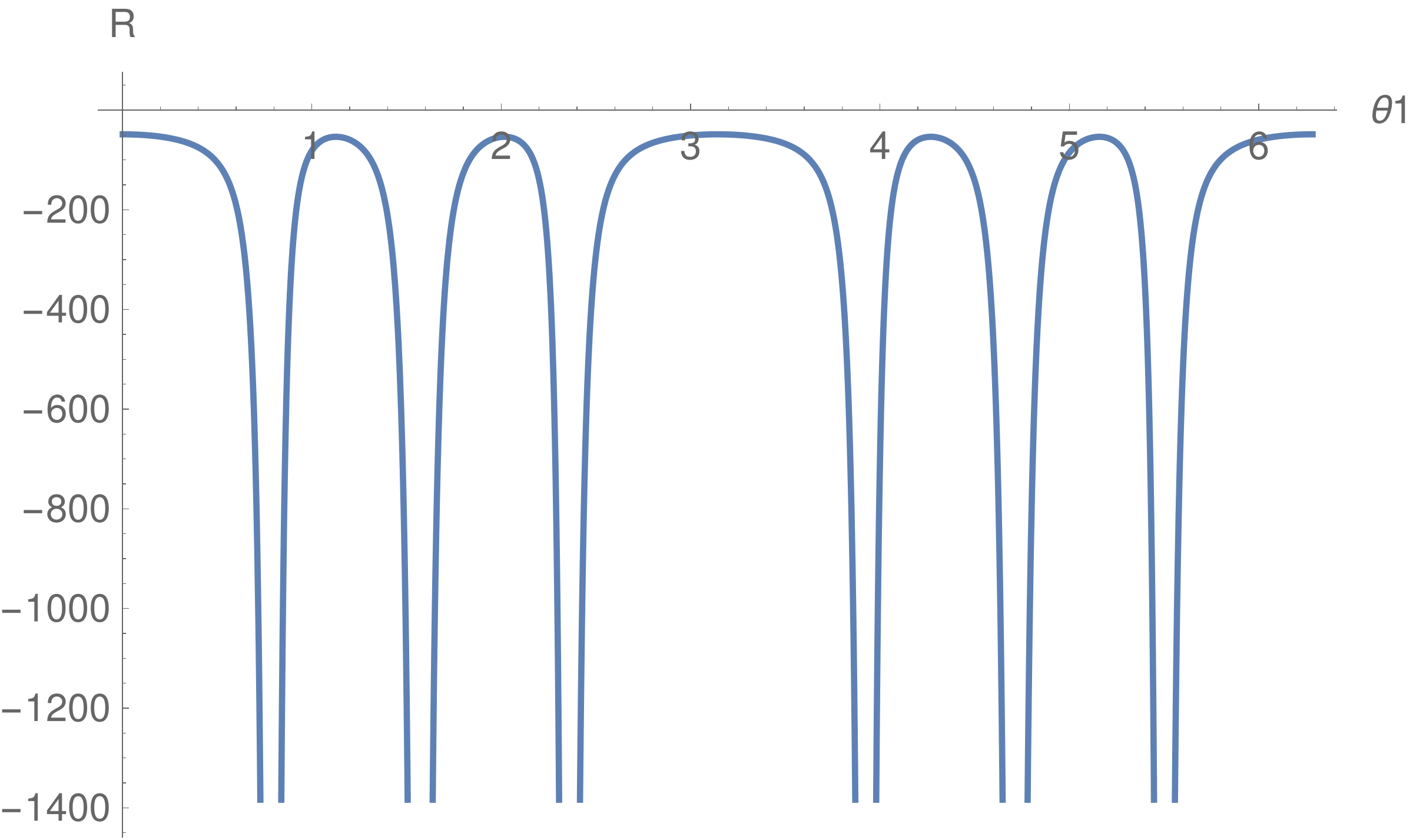}
\caption{ \label{fig:R_CP2_limit}
Scalar curvature \eqref{eq:Ricci_scalar_CP2_infiniteP} for the unitary space $\mathrm{SU}(3)$ in the limiting 
case of penalty $P \rightarrow \infty$ applied to all the generators of the maximal subgroup.}
\end{figure*}

We consider the limit when $P \rightarrow \infty,$ which means that the motion along
 the subgroup directions is strongly penalised.
In this limit the Ricci scalar is
\bea
\lim_{P \rightarrow \infty} R &=&
- \frac{3}{2} \left[  \mathrm{sec} (2 \theta_1) (11 \mathrm{sec} (2 \theta_1) +12)  \right. \nl 
& & \left. + 4 \mathrm{sec}^2 \theta_1 + 5 \right]  \, .
\label{eq:Ricci_scalar_CP2_infiniteP}
\eea
As can be seen in Fig. \ref{fig:R_CP2_limit}, in such a case the Ricci scalar is
 always negative and contains  singularities.
In the opposite limit $P\rightarrow 0$ we instead obtain everywhere a positive
 and divergent Ricci scalar, since it contains a singular term proportional to $P^{-1}.$

The behaviour of the curvature in this example is similar
to the one qubit case with $Q=1$ and $P$ generic,
which was studied in detail in \cite{Brown:2019whu}.


\section{Submersion for $1$ qubit}
\label{1qubit-appe}

Let us apply the method of submersion to the 1 qubit case.
In order to generate a  state specified by the $(\theta, \phi)$ angles 
on the Bloch sphere starting from $|0 \rangle$, 
we can use the following unitary
\beq
U_\theta =\exp \left[  \frac{i \, \theta}{2} \le  \s_x   \cos \phi +  \s_y\sin \phi \ri  \right]\, .
\eeq
The action of unbroken generators can be parametrized by $V_\lambda$
\beq
 V_\l=\exp \le \, i \, \frac{\s_z}{2} \l \ri \, ,
\eeq
and the generic $SU(2)$ transformation is
\bea
U&=&U_\theta V_\l  \nl 
&=&\left(
\begin{array}{cc}
 e^{\frac{i \lambda }{2}} \cos \left(\frac{\theta }{2}\right) & i \sin \left(\frac{\theta }{2}\right)
   e^{-\frac{1}{2} i (\lambda +2 \phi )} \\
 i \sin \left(\frac{\theta }{2}\right) e^{\frac{1}{2} i (\lambda +2 \phi )} & e^{-\frac{i \lambda }{2}} \cos
   \left(\frac{\theta }{2}\right) \\
\end{array}
\right) \, . \nonumber
\eea
The submersion is realised by the projection
 \beq
\pi : \qquad (\l, \theta, \phi) \rightarrow (\theta, \phi) \, ,
\eeq
and the vertical space is spanned by $\p_\l$.

The metric on the unitary space $M$, with penalties $P$ and $Q$ as in section \ref{1qubit-unitaries},  is
\bea
&& ds^2=\frac14 \left\{  (\Tr [i dU \, U^\dagger \, \s_x])^2 \right. \\
&& \left.+ Q \, (\Tr [i dU \, U^\dagger \, \s_y])^2
+P \, (\Tr [i dU \, U^\dagger \, \s_z])^2  \right\} \, ,
\nonumber
\eea
where
\beq
dU=\frac{\p U}{\p \theta} d \theta+\frac{\p U}{\p \phi} d \phi+\frac{\p U}{\p \lambda} d \lambda \, .
\eeq
Explicitly, we find
\beq
dU U^\dagger=i \left(
\begin{array}{cc}
a & b \\
b^* & -a\\
\end{array}
\right) \, .
\eeq
where
\bea
a &=& \frac{1}{2}  ((d\lambda +d \phi) \cos \theta -d \phi) \, , \nl 
b &=&   \frac{1}{2}   e^{-i \phi } (  d \theta - i (d \lambda +d \phi) \sin \theta )  \, .
\eea

Using the unitary metric, we can find the horizontal vectors fields
(which are defined as orthogonal to the vertical direction $\p_\l$)
   \begin{widetext}
\bea
h_1&=&\p_\theta -\frac{(Q-1) \sin \theta  \sin 2 \phi }
{2 \left(P \cos^2 \theta +\sin^2 \theta  \left(Q \cos^2 \phi   +\sin^2 \phi \right)\right)} \p_\l \, , \nl
h_2 &=&\p_\phi+
\frac{-2 P \cos^2 \theta +2 P \cos \theta -\sin^2 \theta   ((Q-1) \cos 2 \phi +Q+1)}
{2 \left(P \cos^2 \theta +\sin^2 \theta  \left(Q \cos^2 \phi   +\sin^2 \phi  \right)\right)} \p_\l \, , 
\eea
   \end{widetext}
which have the property $\pi (h_1)=\p_\theta$, $\pi (h_2)=\p_\phi$.

\begin{figure*}
\begin{center}
\includegraphics[scale=0.45]{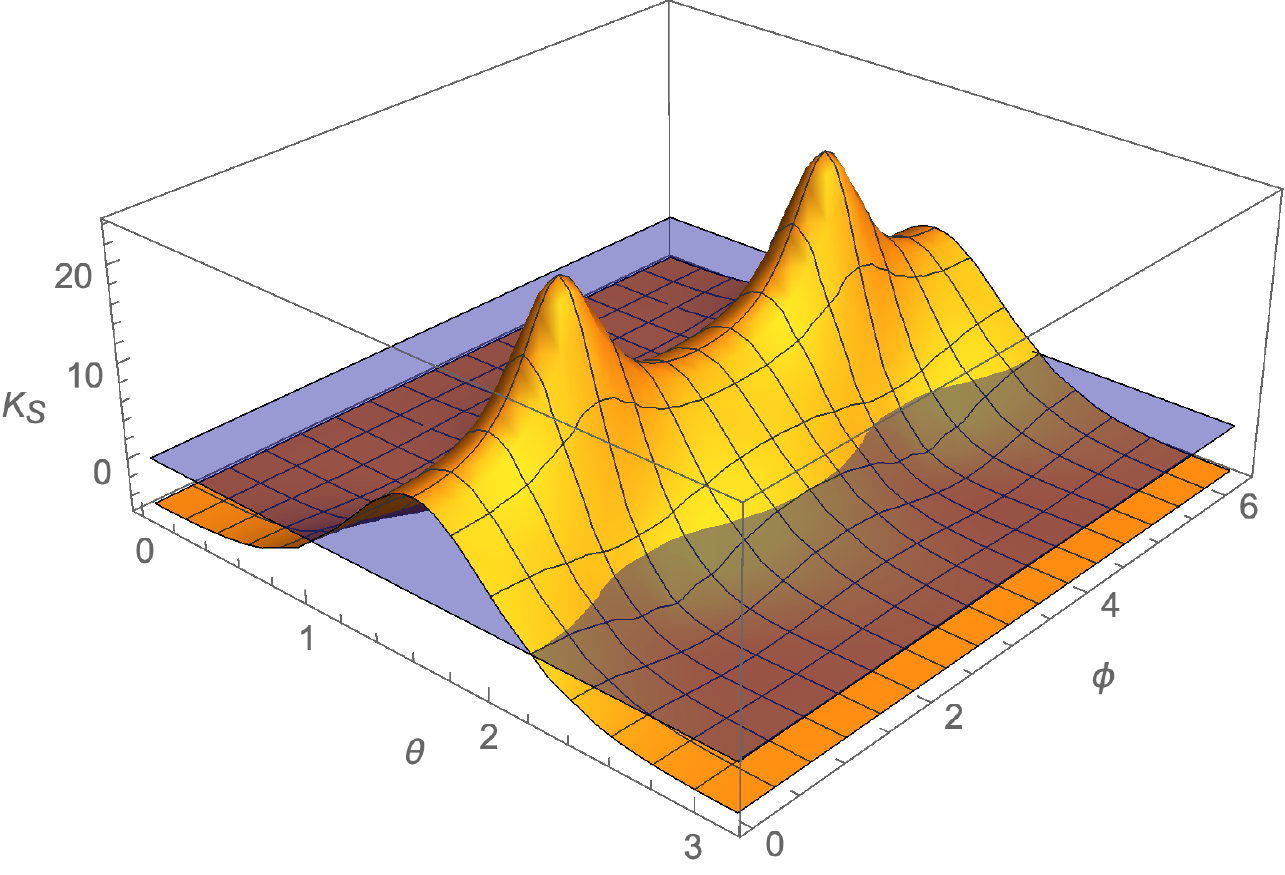} 
\qquad \includegraphics[scale=0.45]{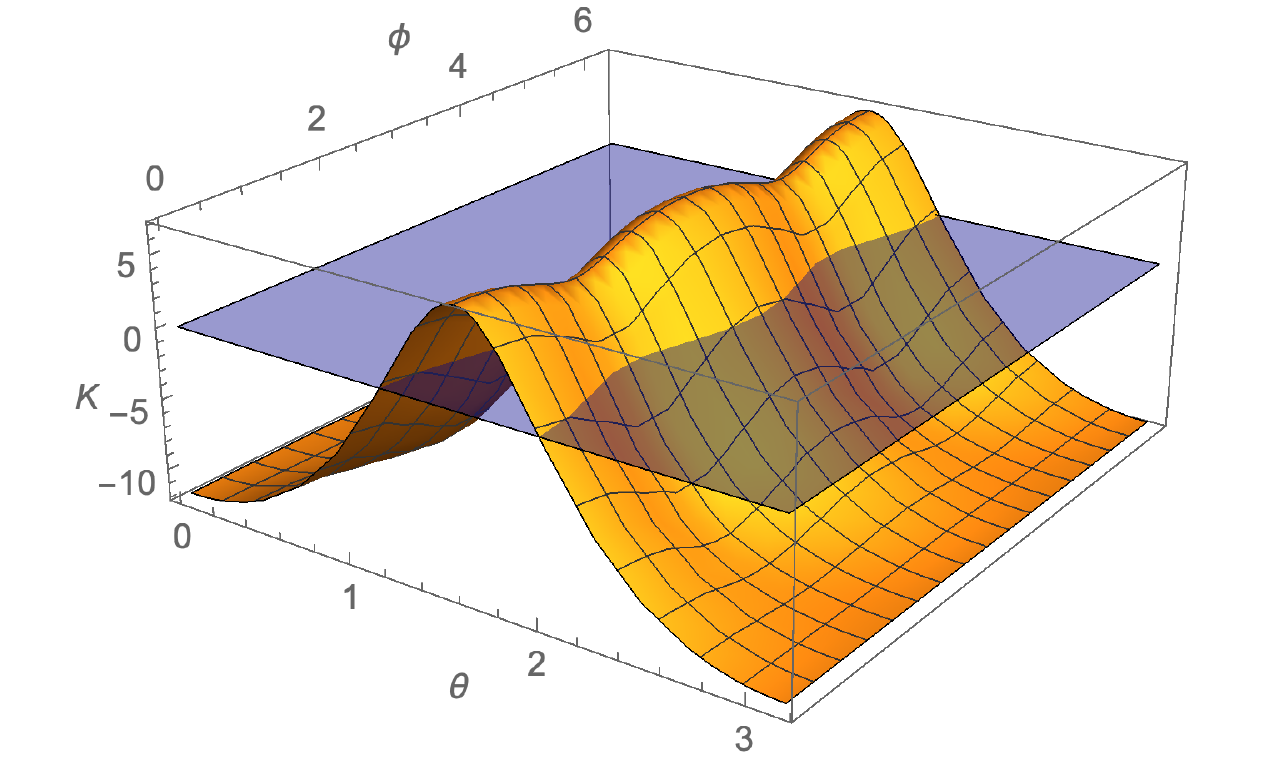}
\qquad   \includegraphics[scale=0.45]{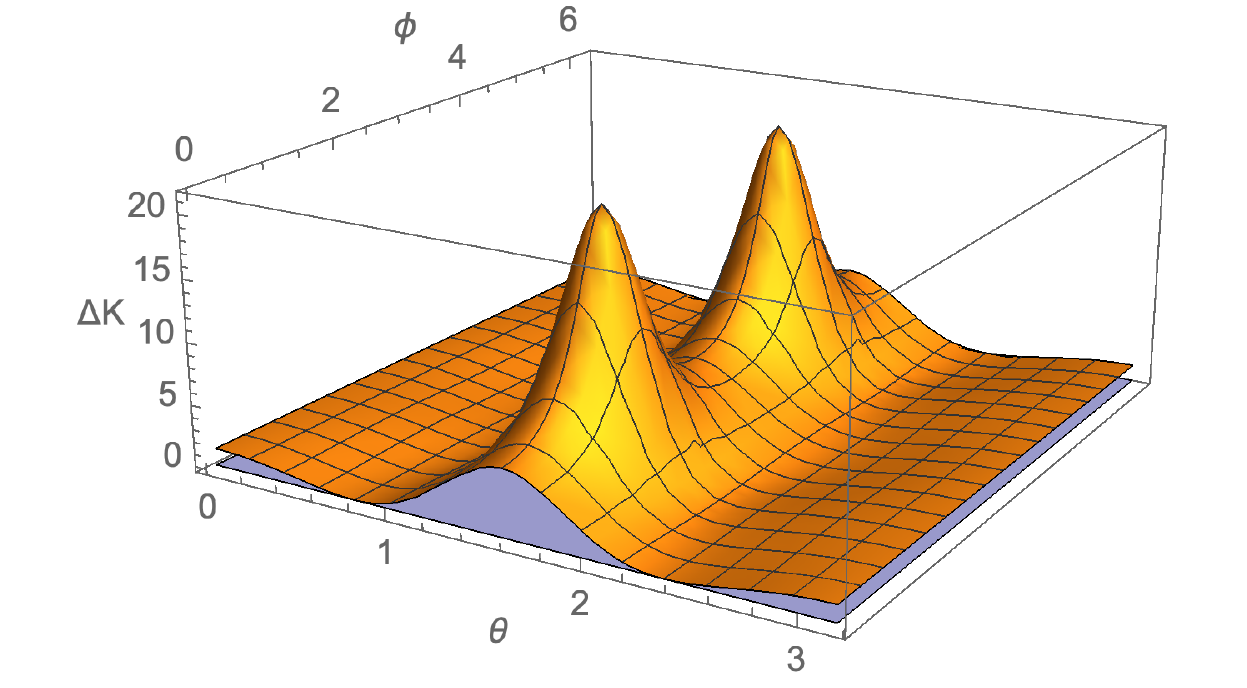}
\caption{  \label{sezionali-oneill}
Comparison of $K_S(\tilde{h}_1, \tilde{h}_2),$  $K(h_1, h_2)$ and $\Delta K$ as a function of $(\theta,\phi)$.
  The numerical values $P=6$, $Q=3$ have been used for illustrative purposes.}
\end{center}
\end{figure*}

Then we can use eq. (\ref{oneill}) to find the curvature in the states space,
using the results for the $1$ qubit unitaries
 in section \ref{1qubit-unitaries}.
An explicit calculation gives the curvature in the states space:
\beq
R=\frac{ \a }{\b} \, ,
\label{scurvature-1qubit}
\eeq
where
\bea
 \a &=& 8 \left\{ - 2  (Q-1)  \sin^2 \theta  \cos^2 \phi  \,\,\,   \times \right. \nl
&& \left. \left[-P^2+(P-1) \cos^2 \theta   (P-Q)^2+P+Q^2\right]    \right. \nl
  &&  \left.  +  (P-1) \cos^2 \theta  \left[-2 \left(P^2-Q^2+Q\right) \right. \right. \nl 
&& \left. \left.  -(P-1)   (Q-1)   (P-Q) \cos^2 \theta \right]  \right. \nl
   &&   \left.    + (P-1) (Q-1)^2  (P-Q)  \sin^4 \theta \cos^4 \phi \right. \nl 
&& \left.   +P (Q-1)+(Q-1) Q +P^2 \right\} \, ,
\eea
\bea
\b & = & P Q \left[ (P-1) \cos^2 \theta \right. \nl
&& \left. +(Q-1) \sin^2 \theta  \cos^2 \phi +1\right]^2 \, ,
   \nonumber
\eea
which matches with the one that can be calculated directly from the states
metric in \cite{Brown:2019whu}.
The difference of the sectional curvatures between the unitary and the state spaces matches with O'Neill formula
\bea
\Delta K &=& K_S(\tilde{h}_1, \tilde{h}_2)- K(h_1, h_2) \nl
& =&  \frac{3}{4}  \frac{| \mathcal{V} ([h_1,h_2]) |^2}{|h_1|^2
|h_2 |^2-\langle h_1, h_2 \rangle^2 } \, .
\eea
The plot of $K_S(\tilde{h}_1, \tilde{h}_2),$  $K(h_1, h_2)$ and $\Delta K$ 
for particular values of the penalties is shown in Fig. \ref{sezionali-oneill}.

\section{Shear tensor equation}
\label{shear-appendix}

The equation for  the traceless part $\sigma_{\a \b}$
 in Euclidean signature is:
\bea
\frac{D \s_{\mu \nu}}{d \lambda}&=&- \frac{2}{d-1} \Theta \s_{\mu \nu}
 -\s^{\s}_{\mu} \s_{\nu \s} \nl 
&+&\frac{1}{d-1} h_{\mu \nu} \s^{\a \b} \s_{\a \b}
- C_{\mu \a \nu \b} u^\a u^\b \nl
&-& \frac{1}{d-2} \bar{R}_{\mu \nu} \, ,
\label{raycha-2}
\eea
where
\beq
\bar{R}_{\mu \nu}=
  h_\mu^\a h_\nu^\b R_{\a \b}
-\frac{1}{(d-1)} R_{\a \b} h^{\a \b} h_{\mu \nu}   
\eeq
is the projected trace-free part of $R_{\mu \nu}$.

The Weyl tensor is given by 
\bea
&& C_{\mu \alpha \nu \beta} = - R_{\mu \alpha \nu \beta} \nl
&&  +\frac{  R_{\mu \beta} g_{\alpha \nu}  
 -   R_{\mu \nu} g_{\alpha \beta} + R_{\alpha \nu} g_{\mu \beta} 
- R_{\alpha \beta} g_{\mu \nu} }{d-2} \nl
&&+ \frac{ g_{\mu \nu} g_{\alpha \beta} - g_{\mu \beta} g_{\alpha \nu}   }{\le d-1 \ri \le d-2 \ri} R 
\eea
and its contraction with the normalized velocity is
\beq
C_{\mu \alpha \nu \beta} \, u^{\alpha} \le \sigma \ri u^{\beta} \le \sigma \ri = \frac{1}{q_{\sigma}} C_{\mu \sigma \nu \sigma} \, .
\eeq
Recalling that in our basis $R_{\mu \sigma \nu \sigma} = - R_{\mu \sigma \sigma \nu} \neq 0$ 
only if $\mu = \nu$ and that both the metric and the Ricci tensor are diagonal, we conclude that $C_{\mu \sigma \nu \sigma} \neq 0$ only if $\mu = \nu$. 
However, if $\mu = \nu = \sigma$, we have $C_{\mu \sigma \nu \sigma} = 0$. 
Therefore, the only relevant non-vanishing components of $C_{\mu \sigma \nu \sigma}$
 are the ones with $\mu = \nu = \rho \neq \sigma$. These components read
\bea
&&C_{\rho \sigma \rho \sigma} 
= q_{\rho} \, q_{\sigma} \left[ K \le \rho \, , \sigma \ri
 - \frac{ R_{\rho} + R_{\sigma}  }{d-2} \right.  \nl 
&& \left.  + \frac{R}{\le d-1 \ri \le d-2 \ri}  \right] \, .
\eea
The only non-vanishing components of the Weyl tensor contraction with the normalized velocity are
the ones with $\rho \neq \sigma$:
\bea
&& C_{\rho \alpha \rho \beta} \, u^{\alpha} \le \sigma \ri u^{\beta} \le \sigma \ri = \\
&& = q_{\rho} \left[ K \le \rho \, , \sigma \ri - \frac{ R_{\rho} + R_{\sigma} }{d-2} + \frac{R}{\le d-1 \ri \le d-2 \ri}  \right]
\nonumber
\eea

A direct calculation gives that $\bar{R}_{\mu \nu}$ is non-vanishing only if $\mu = \nu = \rho \neq \sigma$:
\beq
\bar{R}_{\rho \rho}
= q_{\rho} \left[ R_{\rho} - \frac{1}{d-1} \le R - R_{\sigma} \ri \right] \, .
\eeq
The non-vanishing components of the tensor entering into the shear equation (\ref{raycha-2}) are thus
the ones with $\rho \neq \sigma$:
\bea
 C_{\rho \alpha \rho \beta} \, u^{\alpha} \le \sigma \ri u^{\beta} \le \sigma \ri + \frac{1}{d-2} \bar{R}_{\rho \rho}  \nl
= q_{\rho} \left\lbrace K \le \rho \, , \sigma \ri - \frac{1}{d-1} R_{\sigma} \right\rbrace \, .
\eea
Note that in the one qubit case ($d=3$), by means of eqs. (\ref{sectional-1qubit}) and (\ref{1qubit-Ricci}), all the components of the above tensor vanish for $G_x$ if $P=Q$, for $G_y$ if $P=1$ and for $G_z$ if $Q=1$. In these cases, from eq. (\ref{raycha-2}) we get that if the shear tensor $\sigma_{\a \b}=0$, then it vanishes along all the geodesic.

\end{document}